\documentclass[%
 aip,
 app,
 amsmath,amssymb,
 reprint,%
]{revtex4-1}

\usepackage{graphicx}
\usepackage{dcolumn}
\usepackage{bm}

\usepackage{xcolor}		
\usepackage[colorlinks = true,
            linkcolor = purple,
            urlcolor  = blue,
            citecolor = cyan,
            anchorcolor = black]{hyperref}

\usepackage[utf8]{inputenc}
\usepackage[T1]{fontenc}
\usepackage{mathptmx}
\usepackage{etoolbox}
\usepackage[english]{babel}

\usepackage{booktabs} 		
\usepackage{nicefrac}		
\usepackage{microtype}		
\usepackage{siunitx}
\DeclareSIUnit{\bel}{B}
\DeclareSIUnit{\belmilliwatt}{Bm}
\DeclareSIUnit{\dBm}{\deci\belmilliwatt}
\DeclareSIUnit{\rad}{rad}

\usepackage{bm} 
\usepackage[capitalise]{cleveref}

\newcommand{\iu}{\mathrm{i}\mkern1mu}

\def\app#1#2{%
  \mathrel{%
    \setbox0=\hbox{$#1\sim$}%
    \setbox2=\hbox{%
      \rlap{\hbox{$#1\propto$}}%
      \lower1.1\ht0\box0%
    }%
    \raise0.25\ht2\box2%
  }%
}

\makeatletter
\newcommand*{\balancecolsandclearpage}{%
  \close@column@grid
  \cleardoublepage
  \twocolumngrid
}

\def\@email#1#2{%
 \endgroup
 \patchcmd{\titleblock@produce}
  {\frontmatter@RRAPformat}
  {\frontmatter@RRAPformat{\produce@RRAP{*#1\href{mailto:#2}{#2}}}\frontmatter@RRAPformat}
  {}{}
}%
\makeatother

\begin{document}

\preprint{AIP/123-QED}

\title[DSP Techniques for Noise Characterization of Lasers and Optical Frequency Combs]{Digital Signal Processing Techniques for Noise Characterization of Lasers and Optical Frequency Combs: A Tutorial}
\author{Jasper Riebesehl}
\author{Holger R. Heebøll}
\author{Aleksandr Razumov}
\author{Michael Galili}%
\author{Darko Zibar}
\email{dazi@dtu.dk}
\affiliation{ 
Department of Electrical and Photonics Engineering, Technical University of Denmark, DK-2800, Kgs. Lyngby, Denmark
}
\date{\today}

\begin{abstract}
Performing noise characterizations of lasers and optical frequency combs on sampled data offers numerous advantages compared to analog measurement techniques.
One of the main advantages is that the measurement setup is greatly simplified. Only a balanced detector followed by an analog-to-digital converter is needed, allowing all the complexity to be moved to the digital domain.
Secondly, near-optimal phase estimators are efficiently implementable, providing accurate phase noise estimation in the presence of the measurement noise. Finally, joint processing of multiple comb lines is feasible, enabling computation of phase noise correlation matrix, which includes all information about the phase noise of the optical frequency comb. This tutorial introduces a framework based on digital signal processing for phase noise characterization of lasers and optical frequency combs. The framework is based on the extended Kalman filter (EKF) and automatic differentiation. The EKF is a near-optimal estimator of the optical phase in the presence of measurement noise, making it very suitable for phase noise measurements. Automatic differentiation is key to efficiently optimizing many parameters entering the EKF framework. 
More specifically, the combination of EKF and automatic differentiation enables the efficient optimization of phase noise measurement for optical frequency combs with arbitrarily complex noise dynamics that may include many free parameters.
We show the framework's efficacy through simulations and experimental data, showcasing its application across various comb types and in dual-comb measurements, highlighting its accuracy and versatility. Finally, we discuss its capability for digital phase noise compensation, which is highly relevant to free-running dual-comb spectroscopy applications.
\end{abstract}

\maketitle

\section{Introduction}

Optical frequency combs are indispensable tools in many scientific and engineering disciplines \cite{diddams_optical_2020, fortier_20_2019}. Over the past two decades, the field of optical frequency combs has witnessed significant advancements, leading to the development of various types of frequency combs with distinct characteristics and applications \cite{newbury_low-noise_2007, tian_noise_2021, kippenberg_microresonator-based_2011, haus_noise_1993}.
The development of broad and stable combs has led to their integration in various fields, notably playing vital roles in the establishment of ultra-stable frequency references for metrology applications, high-capacity coherent telecommunication, and high-precision spectroscopy\cite{lundberg_frequency_2018, picque_frequency_2019, schioppo_comparing_2022, nardelli_10_2022}.

While extensive research efforts have been made to realize various types of optical frequency combs, the development of novel noise characterization techniques has not followed a similar trend. Performing noise characterization of optical frequency combs poses a challenge due to their complex phase noise dynamics and the corresponding scaling with comb-line number \cite{razumov_subspace_2023,heeboll_subspace_2024,razumov_phase_2024,paschotta_noise_2004-1, paschotta_noise_2004}. Additionally, different comb lines may have different signal-to-noise ratios (SNRs), requiring measurement techniques to estimate phase noise over a wide range of SNRs accurately \cite{brajato_bayesian_2020}.

Understanding and characterizing noise properties of optical frequency combs is paramount to harnessing their full potential. The magnitude of their phase noise is a crucial property as its presence degrades the frequency stability of the individual comb lines, resulting in the limited system performance \cite{lundberg_phase-coherent_2020, diddams_optical_2020}.

Various phase noise measurement techniques have been proposed in the literature \cite{ nishimoto_investigation_2020, ishizawa_phase-noise_2013, xie_photonic_2017}. As these methods are implemented in the analog domain, they introduce additional complexity to the experimental setup.
More importantly, analog methods usually rely on optical bandpass filters to isolate single comb lines from the spectrum. In the aforementioned approaches, the measurement accuracy can be limited by the SNR of single comb lines. This is because the combination of low-power comb lines and measurement noise may lead to low SNR, leading to an inaccurate phase noise estimate. While cross-correlation methods can reduce the impact of measurement noise on the phase noise estimate, they require additional calibration and introduce complexity \cite{yuan_correlated_2022}. Additionally, optically filtering out single lines can be infeasible for very narrowly spaced combs.

Another drawback of performing phase noise characterization of frequency combs based on line-by-line measurement is the inability to extract the time-domain correlations between neighboring lines. To remedy this, analog measurement methods have been developed to measure multiple comb lines simultaneously \cite{ansquer_unveiling_2021,lao_quantum_2023}. However, these methods again introduce more experimental complexity while still being limited to measuring only a few lines.

The complementary approach of using balanced detection in combination with digital signal processing for phase noise estimation shifts most of the measurement complexity into the digital domain. The availability of ultra-wideband photodetectors and analog-to-digital (ADC) converters at the order of \SI{100}{\giga\hertz} bandwidth allows the detection of many comb lines simultaneously in multi-heterodyne measurements. The sampled data is stored offline, allowing advanced digital signal processing (DSP) for joint phase noise estimation of multiple frequency comb lines to be efficiently implemented \cite{lundberg_frequency_2018, vedala_phase_2017}.

Applying DSP techniques thereby allows a simultaneous phase noise characterization of multiple comb lines while retaining a minimal measurement setup \cite{schlatter_simultaneous_2007}. The extracted phase noise can then be used to compute the phase noise correlation matrix, which can then be used to identify the underlying phase noise sources and their comb line number dependent scaling \cite{brajato_bayesian_2020, razumov_subspace_2023}.

However, some of the aforementioned DSP-based phase noise characterization methods employ digital bandpass filtering of individual comb lines, thereby inheriting some of the limitations of analog techniques. This implies that the SNR of the individual lines limits the estimation accuracy of each comb line's phase noise, and narrowly spaced combs still pose an issue.

Bayesian filtering implemented as an extended Kalman filter offers joint digital signal processing of multiple comb lines. It eliminates the shortcomings of digitally filtering individual comb lines while maintaining the reduced experimental complexity. In particular, the EKF can provide a near-optimal phase noise estimation in the presence of measurement noise \cite{zibar_approaching_2021}. Given an analytical description (model) of the detected signal and the associated phase noise sources, the EKF can distinguish between phase and measurement noise. This distinction allows the digital suppression of the measurement noise and an improved phase noise estimate over the state-of-the-art conventional methods \cite{brajato_bayesian_2020}.
The EKF inherently provides confidence intervals (CIs) with the phase noise estimates, allowing the estimate's uncertainty to be quantified. In a laser phase noise characterization, the EKF was able to predict phase noise up to \SI{7.6}{\deci\bel} below the measurement noise floor with \SI{95}{\percent} confidence \cite{riebesehl_quantifying_2024}.

In \cite{brajato_bayesian_2020}, a framework based on the EKF is presented to characterize the phase noise of an electro-optic (EO) comb. The EKF operates in the time domain using a joint model of the detected comb lines, so no prior bandpass filtering is required. The estimation accuracy is improved compared to the conventional DSP methods, and the phase noise correlations can be recovered.
In this framework, however, the computational complexity scaling of the EKF is $\mathcal{O}(N^3)$ where $N$ is the number of detected comb lines. This scaling is unfavorable and practically disqualifies the method for combs with many lines.

In \cite{burghoff_generalized_2019}, the EKF was used to compensate for phase noise in a dual-comb spectroscopy setup. Here, the complexity scaling issue was avoided by estimating the carrier-envelope frequency and the repetition rate phase noise of the comb instead of estimating each line. The complete reconstruction of a noise-free comb spectrum from a free-running comb is demonstrated. However, this method is tailored toward applying noise compensation in dual-comb spectroscopy rather than characterization.

This tutorial introduces an improved EKF-based framework for phase noise characterization of optical frequency combs.
The framework's significant improvement compared to \cite{burghoff_generalized_2019} is the inclusion of automatic differentiation (AD). As the model within the framework may contain many free parameters that need to be determined, AD allows for gradient-based static parameter optimization. This approach to model parameter optimization significantly reduces the complexity and enables efficient optimization in high-dimensional search spaces.
Moreover, AD allows for the introduction of a latent phase noise space, a compressed representation of comb line phase noise. The structure of the latent space is learned directly from the measurement data. The introduction of an optimizable latent space not only solves the complexity scaling problem. It also enables the search for phase noise modes beyond the commonly agreed upon noise sources in combs \cite{razumov_subspace_2023}.

We apply AD and adaptive gradient-based optimization to learn the structure of the latent space. AD allows the optimization of an analytical model that describes the signal. Critically, AD enables the calculation of all model parameter gradients within any differentiable EKF model expression, eliminating the need for almost all auxiliary parameter estimation techniques.
In this joint optimization process, the models of the measured signal and the statistical properties of the noise sources are learned.
This versatility makes the method very general and, in principle, applicable to any comb.

In this work, we demonstrate in detail how to use this framework on different types of combs in simulation and experimental data.
The remainder of this tutorial is structured as follows: In section \ref{sec:02_Methodology}, we introduce the required experimental setup and derive analytical expressions to describe the detected signal.
We supply the necessary theoretical background for applying the EKF and optimizing a model. The framework is applied to a synthetic laser beat note measurement to detail every step in the process. 
Additionally, the state-of-the-art conventional phase estimation method is explained and applied.
In section \ref{sec:03_downconverted_combs}, the framework is applied to an EO comb on simulated and experimental data. Here, we demonstrate the versatility of the optimization process, the accuracy of the phase noise estimation, and the extraction of phase noise correlations.
In section \ref{sec:04_dual_comb}, we show how the framework can be used in dual-comb measurements. Using two mode-locked lasers, we present the phase noise estimation process on an experimental measurement. Further, we discuss how this framework can be applied for digital noise compensation similar to the method presented in \cite{burghoff_generalized_2019}.
Finally, we give some concluding remarks and a brief outlook in section \ref{sec:conclusion}.

\section{Methodology} \label{sec:02_Methodology}

\begin{figure*}
\centering
\includegraphics[width=0.9\linewidth]{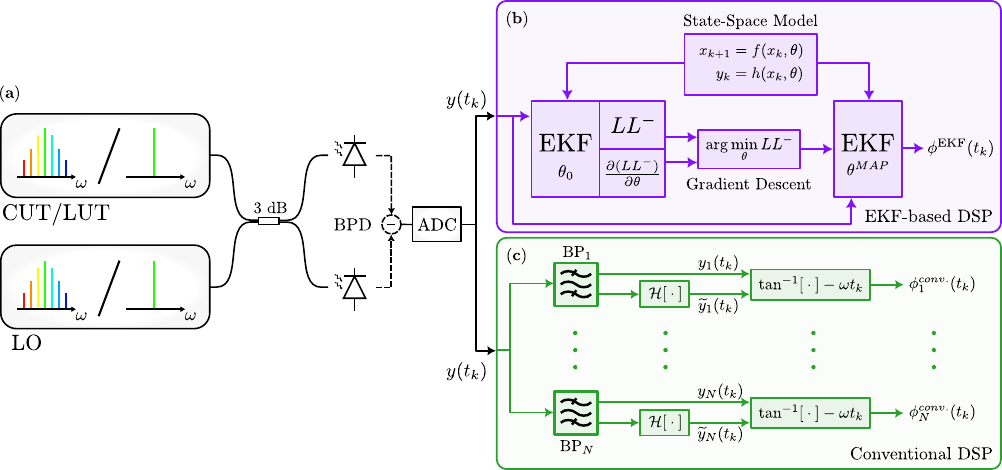}
\caption{Schematic representation of the multi-heterodyne setup as well as the DSP frameworks.
(a) Experimental setup to generate multi-heterodyne beat note signals from a comb/laser under test (CUT/LUT) and a local oscillator (LO). BPD: balanced photodetector, ADC: analog-to-digital converter.
(b) Schematic representation of the EKF-based phase noise estimation framework, which is discussed in \cref{sec:EKF_method}.
(c) Schematic representation of the conventional phase noise estimation method, which is discussed in \cref{sec:hilbert-method}.
}
\label{fig:setup}
\end{figure*}

This section describes the complete framework in detail.
First, the experimental setup common to all multi-heterodyne measurements considered in this work is described.
A conventional phase noise measurement method using simple digital phase noise estimators is briefly explained.
Finally, the EKF-based method is detailed. We describe how AD is applied in an efficient and versatile optimization framework. 
At the end of the section, we step through the example of a laser phase noise characterization as it is conceptually simpler than the frequency comb case while still showcasing the intricacies of the method.

\subsection{Setup for numerical simulations and experimental measurements}
\label{sec:setup}
The experimental setup described in this work is shown in \cref{fig:setup}(a). The same setup is also used for simulation studies.
It is used to generate heterodyne beat note signals for use in frequency comb and laser characterizations. The setup requires minimal analog components, including an optical coupler, a BPD, and an ADC, making it easy to reproduce.

Throughout this article, the same experimental setup is used for all characterization studies, whether in simulation or experiment. All optical components in the setup are assumed to be fiber-bound, but a free-space setup would also work.

In the measurement process, the comb or laser under test (CUT/LUT) and the local oscillator (LO) are combined using a balanced fiber coupler. The combined signal is then detected by a balanced photodetector, which suppresses intensity noise and reduces shot noise contributions, resulting in an improved signal-to-noise ratio (SNR) of the signal \cite{agrawal_fiberoptic_2010}.

The detection process produces one or multiple beat notes, depending on the number of optical tones in the light sources and their detuning with respect to each other. This setup is known as (multi-) heterodyne detection.

To find a general analytical description of the detected signal, we assume that both inputs to the coupler are frequency combs. We define each optical comb as a sum of optical lines that are defined by a time-dependent amplitude, a frequency, and a time-dependent phase noise term. The frequency of each line is composed of a fixed offset from zero, often referred to as the carrier-envelope offset (CEO) frequency $\omega_{ceo}$, and an integer multiple of a repetition frequency $\omega_{rep}$ that defines the spacing of the comb lines. The complex electric fields for both sources can be expressed as
\begin{subequations}
    \begin{align}
        E_{CUT}(t) &= \sum_{l=0}^\infty \sqrt{P^{CUT}_{l} + \Delta P^{CUT}_{l}(t)} \nonumber\\
        &\qquad \times e^{\iu((\omega^{CUT}_{ceo}+l\omega^{CUT}_{rep}) t) + \phi^{SUT}_{l}(t))} \label{eq:E_field_SUT}\\
        E_{LO}(t) &= \sum_{l=0}^\infty \sqrt{P^{LO}_{l} + \Delta P^{LO}_{l}(t)} \nonumber\\
        &\qquad \times e^{\iu((\omega^{LO}_{ceo}+l\omega^{LO}_{rep}) t) + \phi^{LO}_{l}(t))} \label{eq:E_field_RS}
    \end{align}
    \label{eq:E_field_general}
\end{subequations}
where $P_{l} + \Delta P_{l}(t)$ is the power of the l-th line plus its time-dependent fluctuation, and $\phi_{l}(t)$ is the time-dependent phase noise of the l-th line.
As the infinite sums suggest, an optical frequency comb may have arbitrarily many comb lines. In practice, however, only a subset of these lines is of interest. Often, experimental thresholds like a minimal detectable optical power or the optical frequency of given lines, limit which lines are detectable.

This general description can model any combination of CUT/LUT and LO. Mathematically, a LUT is equivalent to a CUT with only a single line.
As this is a very general signal description, it can be unwieldy and impractical. We will, therefore, examine multiple special cases of this expression throughout this article and detail how this expression simplifies.

Propagating the electric fields through a balanced \SI{3}{\deci\bel} fiber coupler causes a superposition of the input fields at its outputs, defined as
 \begin{align}
    \begin{bmatrix}
       E_{a}(t) \\
       E_{b}(t) \\
     \end{bmatrix}
         = \frac{1}{\sqrt{2}}
     \begin{bmatrix}
       1 & \iu \\
       \iu & 1 \\
     \end{bmatrix}
    \cdot
     \begin{bmatrix}
       E_{CUT}(t) \\
       E_{LO}(t) \\
     \end{bmatrix}. \label{eq:optical_coupling}
\end{align}
Finally, detection in the BPD and subsequent digitization by the ADC will lead to the detected signal
\begin{equation}
    \begin{split}
    y(t_k) &= \eta \left(|E_{b}(t_k)|^2 - |E_{a}(t_k)|^2\right) \\
    &= 2\eta \sum_{l=0}^\infty \sum_{m=0}^\infty
    \sqrt{(P^{CUT}_l + \Delta P^{CUT}_l(t_k))(P^{LO}_m + \Delta P^{LO}_m(t_k))}  \\
    &\quad \times \text{sin}[(\omega^{CUT}_{ceo} - \omega^{LO}_{ceo} + l\omega^{CUT}_{rep} - m\omega^{LO}_{rep})t_k \\
    &\qquad  + \phi^{CUT}_l(t_k) - \phi^{LO}_m(t_k)] + \xi(t_k). \label{eq:BPD_with_RIN}
    \end{split}
\end{equation}
Here $t_k=k\Delta T_s$ indicates time discretization by the ADC with sampling period $\Delta T_s$, $\eta$ is a detection proportionality constant, and $\xi(t_k)$ is detector measurement noise.
$\xi(t_k)$ contains contributions originating from thermo-electric noise in the detector, shot noise, and ADC quantization noise \cite{agrawal_fiberoptic_2010, schlatter_simultaneous_2007}.
While bandwidth limitations in the detection system do not allow for an infinite number of detected comb lines, we keep the notation for simplicity.

If a single photodiode were used instead of a BPD, the detected signal would be proportional to $|E_{a}|^2$. This would result in constant power offset terms and no amplitude noise suppression. Additionally, single photodiode detection has a \SI{3}{\deci\bel} SNR penalty compared to BPD.
While BPD is the superior detection method, single PD detection can be used to the same effect if amplitude noise and SNR are not a concern.

Using a BPD, the power fluctuations $\Delta P_{l/m}(t_k)$ enter Eq. \eqref{eq:BPD_with_RIN} only in square root order. In addition they typically are significantly less prominent compared to phase noise, such that they can often be neglected. Under this assumption, the expression simplifies to
\begin{equation}
    \begin{split}
    y(t_k) &\approx 2\eta \sum_{l=0}^\infty \sum_{m=0}^\infty \sqrt{P^{CUT}_l P^{LO}_m} \\
    &\quad \times \text{sin}[(\omega^{CUT}_{ceo} - \omega^{LO}_{ceo} + l\omega^{CUT}_{rep} - m\omega^{LO}_{rep})t_k \\
    &\qquad  + \phi^{CUT}_l(t_k) - \phi^{LO}_m(t_k)] + \xi(t_k) \\
    &= \sum_{l=0}^\infty \sum_{m=0}^\infty A_{lm} \text{sin}[\omega_{lm} t_k + \phi^{total}_{lm}(t_k)] + \xi(t_k)
    \end{split}
    \label{eq:BPD_no_rin}
\end{equation}
where we have defined $A_{lm} = 2\eta \sqrt{P^{CUT}_l P^{LO}_m}$ as the effective detected amplitudes of the beat notes, $\omega_{lm}=\omega^{CUT}_{ceo} - \omega^{LO}_{ceo} + l\omega^{CUT}_{rep} - m\omega^{LO}_{rep}$ as their effective frequencies, and $\phi^{total}_{lm}(t_k) = \phi^{CUT}_l(t_k) - \phi^{LO}_m(t_k)$ as the total phase noise per beat note.
While we will use this simplification throughout this work, it might not generally apply in all situations, especially when using single PD detection. If significant amplitude fluctuations are present, the conditions for the simplification are not met. This can lead to a biased phase noise estimation in the DSP step of the framework. This assumption should, therefore, be checked thoroughly before using the framework.

The quantities of interest for the phase noise characterization of the CUT are $\phi^{CUT}_l(t_k)$.
However as both $\phi^{CUT}_l(t_k)$ and $\phi^{LO}_m(t_k)$ are randomly fluctuating phase noise terms, it is not possible to separate their difference $\phi^{total}_{lm}(t_k)$. As we can not rely on additional prior knowledge about their statistics, they become indistinguishable.
To still be able to estimate $\phi^{CUT}_l(t_k)$, one of two approaches should be chosen.
For the first option, a LO needs to be selected whose phase noise is negligible compared to the CUT, in which case $\phi^{total}_{lm}(t_k) \approx \phi^{CUT}_l(t_k)$.
The other option is to use a reference with close to identical phase noise properties as the CUT.
It still is impossible to separate their contributions in the time domain, as the time evolution of their phases is uncorrelated.
However, both noise contributions will have the same statistical behavior, which leads to them having approximately identical phase noise power spectral densities (PSDs) $S^{CUT,l}_{\phi}(f) \approx S^{LO,l}_{\phi}(f)$ per line \textit{l}.
As the total phase noise $\phi^{total}_{lm}(t_k)$ is the difference between the individual noise sources, the total PSD can be expressed as $S^{total,ll}_{\phi}(f) = S^{CUT,l}_{\phi}(f) + S^{LO,l}_{\phi}(f) \approx 2 S^{CUT,l}_{\phi}(f)$. Here it is immediately clear that in this specific case, a division by two yields the PSD of the \textit{l}-th line of the CUT. Notably, this argument is only effective when investigating the beat notes that originate from comb lines with the same properties, as indicated with identical line index \textit{l}.
This scenario is useful when multiple copies of the CUT are available.

If neither of these options can be chosen due to a lack of suitable references, the measured phase noise will contain an unknown mixture of both sources. While this is not ideal for the characterization of the CUT, extracting the total phase noise per comb line can still be useful in certain applications.
In dual-comb spectroscopy, for example, the amount of total phase noise per detected comb line can be a limitation \cite{coddington_dual-comb_2016}.
Here the exact composition of $\phi^{total}_{lm}(t_k)$ from the CUT and LO is less important in practice.

\label{sec:hilbert-method}
\subsection{Conventional phase noise estimation algorithm}
In this section, a brief summary of the conventional digital phase noise estimation algorithm is given \cite{brajato_bayesian_2020, zibar_approaching_2021, razumov_subspace_2023}.
It is based on the discrete Hilbert transform $\mathcal{H}$ to obtain the second quadrature of the detected multi-heterodyne beat, followed by an inverse tangent operation to obtain the phase.
A schematic illustration of the algorithm is given in \cref{fig:setup}(c).

Starting from the detected multi-heterodyne signal as in Eq. \eqref{eq:BPD_no_rin}, each beat note needs to be isolated. This is accomplished using a bank of digital bandpass filters. Each filter is centered around the beat note frequencies $\omega_{lm}$. The bandwidth is determined by the frequency difference between the detected notes. 
Applying the filter bank produces a time-domain trace $y_{lm}(t_k)$ for each beat note.
Each $y_{lm}(t_k)$ now only contains one beat note and represents one of the quadratures of the signal.

The Hilbert transform is then used to calculate the orthogonal quadrature $\widetilde{y}_{lm}(t_k)=\mathcal{H}(y_{lm}(t_k))$ by imparting a $\pi/2$ phase shift.
The phase noise $\phi^{total}_{lm}(t_k)$ of each beat note can now be estimated using 
\begin{align}
    \hat{\phi}^{conv.}_{lm}(t_k) = \tan^{-1}\left[y_{lm}(t_k) / \widetilde{y}_{lm}(t_k)\right] - \omega_{lm} t_k.
\end{align}
The term $\omega_{lm} t_k$ represents a linear detrending operation. 
The approach is comparatively low in complexity and does not require any prior knowledge about the signal.

Notably, this method does not attempt to remove measurement noise present in the signal. The measurement noise term $\xi(t_k)$ in Eq. \eqref{eq:BPD_no_rin} is still present in each $y_{lm}(t_k)$ after the bandpass filtering operation.
In high SNR regimes where measurement noise is negligible, this method has been shown to be an optimal estimator of $\phi^{total}_{lm}(t_k)$ \cite{zibar_approaching_2021}.
In moderate to low SNR regimes, however, measurement noise can dominate such that the estimator is no longer optimal. The resulting phase noise estimate then contains measurement noise contributions, as the estimator can not distinguish between the two. The resulting phase noise estimate is biased and overestimates the phase noise magnitude.
In regimes with low SNR and low phase noise, the method can even fail completely to produce a reasonable estimate.

Additionally, the bandpass filtering operation limits the maximum offset frequency at which the phase noise can be estimated. This frequency is half of the bandpass filter's bandwidth, which is half the frequency distance of a given beat note to its neighboring beat note.
This limitation becomes relevant for narrowly spaced combs: In dual-comb spectroscopy, the multi-heterodyne beat notes can have spacings of less than \SI{100}{\hertz}. This method would therefore only allow a phase noise characterization of up to \SI{50}{\hertz}.

Throughout this work, this method will be called the conventional method and serve as a benchmark for the EKF framework.

\subsection{EKF-based phase noise estimation framework}
\label{sec:EKF_method}
In signals with low SNR or low phase noise beat notes, a method that can reduce the impact of measurement noise and isolate phase noise is necessary.
For this purpose, Bayesian filtering in the form of the \textit{extended Kalman filter} (EKF) has been proposed. The EKF has been shown to provide near-optimal phase noise estimates in (multi-) heterodyne measurements in the presence of measurement noise \cite{zibar_approaching_2021, brajato_bayesian_2020}. After formulating a \textit{state-space model} that describes the detected signal in Eq. \eqref{eq:BPD_no_rin}, it can be used to track dynamic quantities inside the model over time that can not be measured directly. These are the phase noise terms and are often called the "hidden" state of the state space model. Simultaneously, the EKF attempts to isolate the measurement noise present in the signal based on its statistical properties. By separating the measurement noise from the phase noise, an accurate phase noise estimate can be produced even in low SNR regimes.

If the analytic expression underlying the detected signal was linear, the regular Kalman filter can be shown to provide the optimal estimate with regards to the mean squared error \cite{sarkka_bayesian_2013}. In our case, however, the expression in Eq. \eqref{eq:BPD_no_rin} is highly non-linear. This is known as the non-linear filtering problem, which does not have an analytical solution. This implies that no optimal filter exists for our purpose. The EKF provides an approximate solution by locally linearizing the state-space model. While the optimality property is lost, this first-order approximation is typically sufficient to produce accurate estimates. We therefore call the estimate "near-optimal".

The EKF operates directly on the detected signal in the time domain given in Eq. \eqref{eq:BPD_no_rin}. All beat notes in the signal are simultaneously part of the state-space model, which makes this a joint estimation approach. This means that no prior bandpass filtering around the individual combines is required as joint estimation of phase noise of each beat note line is performed. Consequently the method does not exhibit the same limitations induced by bandpass filtering as the analog methods or the conventional method.

In the following subsections, the general theory of the EKF is presented. A schematic representation of the algorithm is presented in \cref{fig:setup}(b).
After a rundown of the prerequisites, the EKF is applied to characterize the phase noise of a laser in \cref{sec:application_to_laser}.

\subsubsection{State space model}
\label{sec:state_space_model}
The first step in applying the EKF is the definition of a state space model.
The model consists of a \textit{state-transition function} $\mathbf{f}$ and a \textit{measurement function} $\mathbf{h}$, which have the form
\begin{align}
    \mathbf{x}_{k+1} &= \mathbf{f}(\mathbf{x}_k, \pmb{\theta}) + \mathbf{q}_k \,&\, \mathbf{q}_k \sim \mathcal{N}(0, \mathbf{Q}) \label{eq:transition_eq} \\
    \mathbf{y}_{k} &= \mathbf{h}(\mathbf{x}_k, \pmb{\theta}) + \mathbf{r}_k \,&\, \mathbf{r}_k \sim \mathcal{N}(0, \mathbf{R}) \label{eq:measurement_eq}
\end{align}
where bold lowercase symbols represent vector-valued quantities, and uppercase bold symbols represent matrices.
$\mathbf{y}_{k}$ is the measured signal at time $t_k$, which is modeled by $\mathbf{h}(\mathbf{x}_k, \pmb{\theta})$. In multi-heterodyne measurements, $\mathbf{h}$ is described by the analytical form in Eq. \eqref{eq:BPD_no_rin}. 
The arguments of $\mathbf{h}$ are the vector $\mathbf{x}_k$ which contains values of the hidden variables at sample $k$. $\mathbf{x}_k$ corresponds to $\phi^{total}_{lm}(t_k)$ in Eq. \eqref{eq:BPD_no_rin}.
The second argument $\pmb{\theta}$ is a vector of time-independent parameters of the model, which we will refer to as static parameters. Here, $\pmb{\theta}$ contains all other quantities in Eq. \eqref{eq:BPD_no_rin} that are not phase noise, like the beat note amplitudes and their frequencies.
The additional term $\mathbf{r}_k$ describes an additive measurement noise contribution on the measured signal, which corresponds to the detector noise $\xi(t_k)$.
$\mathbf{r}_k$ is a stochastic variable that is distributed according to a zero-mean Gaussian probability distribution $\mathcal{N}$ with covariance matrix $\mathbf{R}$.

The state-transition function $\mathbf{f}(\mathbf{x}_k, \pmb{\theta})$ models how the hidden variables evolve over time. In addition to the deterministic function $\mathbf{f}$, the state space model includes an additive stochastic noise term $\mathbf{q}_k$. This so-called process noise is a random variable distributed as a zero-mean Gaussian with covariance matrix $\mathbf{Q}$. This allows the modeling of hidden variables based on their statistical properties.
In the multi-heterodyne case, Eq. \eqref{eq:transition_eq} models the phase noise evolution.

This generally requires prior information about the statistical properties of the phase noise.
It has, however, been shown that the phase noise statistics can be approximated \cite{zibar_approaching_2021}. Intuitively, phase noise accumulates random fluctuations over time. Modeling these additive fluctuations to be normally distributed is equivalent to modeling the phase noise evolution as a random walk.
A beat note with phase noise that behaves as a random walk has a Lorentzian lineshape. This is the same lineshape profile a quantum fluctuation limited laser cavity produces, which further motivates this approximation \cite{schawlow_infrared_1958}.
It yields a very simple state-transition function
\begin{align}
    \mathbf{x}_{k+1} &= \mathbf{x}_{k} + \mathbf{q}_k \,&\, \mathbf{q}_k \sim \mathcal{N}(0, \mathbf{Q}) \label{eq:random_walk_model}
\end{align}
which we will use throughout this work.

\subsubsection{Filtering equations}
Once a state space model of the form given in Eqs. (\ref{eq:transition_eq}, \ref{eq:measurement_eq}) that describes the detected signal has been defined, the EKF can be applied.
Applying the EKF consists of iteratively applying the so-called filtering equations.
Given an initial value for the phase noise $\mathbf{x}_0$, applying the EKF the first signal sample $y_1$ produces an estimate for the phase noise $\mathbf{x}_1$ at the first sample. Recursively stepping through the equations produces a vector of phase noise estimates $\mathbf{x}_{1:T}$ for each time step, where $T$ is the number of samples in the signal.

Further, the EKF not only produces the point estimates $\mathbf{x}_{1:T}$. The EKF estimate for each sample is a Gaussian distribution, therefore a vector of covariances $\mathbf{P}_{1:T}$ is produced alongside the point estimates.
For each time step $t_k$, the EKF produces a Gaussian probability distribution $\mathcal{N}(\mathbf{x}_k, \mathbf{P}_k)$ which describe the likelihood for a certain value of the estimate.
This means that the phase noise prediction inherently includes a quantification of the uncertainty of the estimate.
At each $t_k$, the standard deviation of the Gaussian estimate can be interpreted as a confidence interval. 

The filtering equations are given as
\begin{subequations}
\label{eq:EKF_equations}
\begin{align}
  \label{eq:ekf_prediction}
  \hat{\mathbf{x}}_{k} &= \mathbf{f}(\mathbf{x}_{k-1}, \pmb{\theta}) \\
  \hat{\mathbf{P}}_{k} &= J_{\mathbf{f}}(\mathbf{x})\Big|_{\mathbf{x}_{k-1}, \pmb{\theta}} \mathbf{P}_{k-1} J_{\mathbf{f}}(\mathbf{x})^T\Big|_{\mathbf{x}_{k-1}, \pmb{\theta}} + \mathbf{Q} \\
  \mathbf{S}_{k} &= J_{\mathbf{h}}(\mathbf{x})\Big|_{\mathbf{x}_k, \pmb{\theta}} \hat{\mathbf{P}}_{k} J_{\mathbf{h}}(\mathbf{x})^T\Big|_{\mathbf{x}_k, \pmb{\theta}} + \mathbf{R} \\
  \label{eq:ekf_gain}
  \mathbf{K}_{k} &= \hat{\mathbf{P}}_{k} J_{\mathbf{h}}(\mathbf{x})^T\Big|_{\mathbf{x}_k, \pmb{\theta}} \mathbf{S}_{k} ^{-1} \\
  \mathbf{v}_{k} &= \mathbf{y}_{k} - \mathbf{h}(\hat{\mathbf{x}}_{k}, \pmb{\theta}) \\
  \label{eq:ekf_update}
  \mathbf{x}_{k} &= \hat{\mathbf{x}}_{k} + \mathbf{K}_{k} \mathbf{v}_{k} \\
  \mathbf{P}_{k} &= \hat{\mathbf{P}}_{k} - \mathbf{K}_{k} \mathbf{S}_{k} \mathbf{K}_{k}^T
\end{align}
\label{eq:filtering_eqs}
\end{subequations}
where $J_{\mathbf{f}}(\mathbf{x})\Big|_{\mathbf{x}_{k}, \pmb{\theta}}$ indicates the Jacobian of a function $\mathbf{f}$ with respect to $\mathbf{x}$, evaluated at $(\mathbf{x}_{k}, \pmb{\theta})$.

Iterating through these equations and saving the estimates $\mathbf{x}_k$ and their covariances $\mathbf{P}_k$ then produces a series of phase noise estimates, including uncertainties that are based on the presence of measurement noise. 
The equations are given here for completeness; for an in-depth introduction and derivations, see \cite{sarkka_bayesian_2013}.

\subsubsection{Static parameter estimation using automatic differentiation enhanced adaptive optimization} \label{sec:model_optimization}
Thus far we have assumed that the values of the static parameters $\pmb{\theta}$ are known.
This generally is not the case and their values need to be determined. This step is crucial as accurate values for $\pmb{\theta}$ are a strict prerequisite for an accurate phase noise estimation.

In the EKF framework, this is most commonly done by maximizing the posterior distribution $p(\pmb{\theta}|\mathbf{y}_{1:T})$ of $\pmb{\theta}$ given the observations $\mathbf{y}_{1:T}$. It represents the probability of a specific value of $\pmb{\theta}$ being responsible for producing the observations $\mathbf{y}_k$. If no prior information about the optimal value of $\pmb{\theta}$, the prior probability distribution $p(\pmb{\theta})$ is constant. According to Bayes's rule, maximizing the posterior distribution then becomes equivalent to maximizing the likelihood $p(\mathbf{y}_{1:T}|\pmb{\theta})$.

Maximizing this quantity allows the estimation of the optimal static parameters of the model for a given measurement. Equivalently, minimizing the negative log-likelihood
\begin{align}
    LL^- = -\log p(\mathbf{y}_{1:T}|\pmb{\theta})
\end{align}
by solving the optimization problem 
\begin{align}
    \hat{\pmb{\theta}}^{MAP} = \arg \min_{\pmb{\theta}} LL^- \label{eq:LL_argmax}
\end{align}
allows for a data-driven optimization method to find the maximum a posteriori (MAP) parameters $\hat{\pmb{\theta}}^{MAP}$.
Working in logarithmic space increases numerical stability as values of $p(\mathbf{y}_{1:T}|\pmb{\theta})$ can become very large.

Using the EKF, $LL^-$ can be approximated with the expression \cite{sarkka_bayesian_2013}
\begin{align}
    LL^-(\mathbf{y}_{1:T}, \pmb{\theta}) \approx \frac{1}{2} \sum_{k=1}^T \log \det 2\pi \mathbf{S}_k + \mathbf{v}_k^T \mathbf{S}_k^{-1} \mathbf{v}_k. \label{eq:LL_EKF}
\end{align}
All quantities can be calculated by iterating through the filtering equations \eqref{eq:filtering_eqs}.
Finding the optimal parameters $\hat{\pmb{\theta}}^{MAP}$ in practice, therefore, involves the repeated application of the EKF to the signal $\mathbf{y}_{1:T}$.

A common technique to solve the optimization problem in Eq. \eqref{eq:LL_argmax} is the Expectation-Maximization algorithm. It is an iterative optimization scheme in which convergence to a (local) optimum is theoretically guaranteed.
For certain state-space models including the one used in \cite{brajato_bayesian_2020}, Eq. \eqref{eq:LL_argmax} can even be solved using a closed-form analytic expression \cite{dreano_estimating_2017}.

For state-space models with many static parameters $\pmb{\theta}$, however, an analytic form to solve the optimization problem is not available. For high-dimensional $\pmb{\theta}$, an efficient approach to solve Eq. \eqref{eq:LL_argmax} requires the explicit computation of the gradients $\partial LL^-/\partial \pmb{\theta}$.
While it is possible to calculate the gradients manually, it requires the symbolic propagation of the gradients through the filtering equations \eqref{eq:filtering_eqs}.
This quickly becomes infeasible for state-space models with many parameters $\pmb{\theta}$\cite{brajato_bayesian_2020}.
To alleviate this complication, the use of automatic differentiation (AD) to calculate $\partial LL^-/\partial \pmb{\theta}$ has been proposed \cite{gorad_parameter_2020}.
Using AD allows the accurate and efficient evaluation of partial derivatives of any differentiable function specified in a computer program \cite{baydin2018automatic, minkov_inverse_2020}. It keeps track of all elementary operations inside the function to build a graph of the function composition.
By defining the differentiation rules of these elementary operations, the gradient of a function can be calculated by propagating through the composition graph and recursively applying the chain rule of differentiation.
The key advantage of AD is its ability to differentiate arbitrarily complex functions. Instead of the error-prone derivation of symbolic derivatives, AD handles the computation automatically.

To evaluate $\partial LL^-/\partial \pmb{\theta}$, the AD is propagated through the recursive EKF equations \eqref{eq:filtering_eqs} which are required to evaluate $LL^-$. For our implementation of the EKF framework, we used the \texttt{JAX} library, which provides a fast and easy-to-use implementation of AD \cite{deepmind2020jax}. 

With a way to calculate $\partial LL^-/\partial \pmb{\theta}$, we are able to use any gradient-based optimization method. A robust and efficient optimization algorithm is required which performs well in high-dimensional search spaces, since $\pmb{\theta}$ can potentially have many free parameters. Therefore, we propose the use of modern adaptive gradient-based optimizers, which are commonly used in the optimization of large neural networks. These algorithms are designed for fast convergence in high-dimensional optimization problems.
In particular, we will use the AdaBelief optimizer throughout this article as it converged very fast and accurately in testing \cite{deepmind2020jax, zhuang_adabelief_2020}.

To increase numerical stability and constrain individual parameters to certain ranges of values during the optimization process, it is useful to introduce a transformation function
$\Lambda(\widetilde{\theta}) = \theta$
such that the optimization target becomes $LL^-(\Lambda(\widetilde{\theta}))$.
The AD then propagates through $\Lambda$ such that the parameters that are optimized are $\widetilde{\pmb{\theta}}$ instead of $\pmb{\theta}$.
The new optimization problem then has the form
\begin{align}
\hat{\widetilde{\pmb{\theta}}}^{MAP} = \arg \min_{\widetilde{\pmb{\theta}}} LL^-(\widetilde{\pmb{\theta}})\label{eq:LL_argmax_transformed}
\end{align}

The new parameters $\tilde{\pmb{\theta}}$ that are exposed to the optimizer are unconstrained real numbers.
Choosing appropriate transformation functions $\Lambda$ then allows the actual model parameters $\pmb{\theta}$ to be scaled and constrained.

To illustrate this concept, we explain a possible choice of optimization transform function of an exemplary parameter. 
The variance of the measurement noise $\mathbf{R}$ in the measurement equation \eqref{eq:measurement_eq} of the state-space model needs to be strictly non-negative. In addition, its value can become very small such that numerical accuracy may become an issue.
Both of these issues are avoided if $\Lambda$ is chosen to be the exponential function, which maps the real numbers to non-negative real numbers. On top of that, small values close to zero are mapped to large negative values.
This effectively causes the optimization to be performed in logarithmic space.
$\Lambda$ can be a different transformation for every parameter in $\pmb{\theta}$, as long as it is a differentiable operation.

To reduce the computational load during optimization, a truncated version $LL^-_{train} = LL^-(\mathbf{y}_{1:N_{train}}, \pmb{\theta})$ of the full negative log-likelihood can be used which only considers $N_{train} < T$ samples of the measurement vector $\mathbf{y}_{1:T}$. Under the assumption that the measurement and phase noise are approximately stationary, this truncation is a good approximation of the full negative log-likelihood.
A balance between computational load and accuracy has to be found: Considering too few samples will result in a biased estimate of the model parameters while using a large number of samples will result in unnecessary slow convergence. However, the long-term behavior of the system can not be captured by the likelihood if too few samples are considered.

$LL^-$ generally is a non-convex function, which means that the optimization problem potentially has multiple local minima. The optimization algorithm can, therefore, get stuck in a local minimum and can not guarantee a global optimum. To mitigate this problem, we typically run the optimization multiple times with different initial guesses for the model parameters $\pmb{\theta}$ and choose the model parameters that yield the highest likelihood.

To test whether a good optimum was found, the parameters $\hat{\pmb{\theta}}^{MAP}$ can be tested on a section of the measured signal $\mathbf{y}$ that was not used to calculate $LL^-_{train}$ in the optimization phase.
To test a set of parameters after convergence in the optimization phase, we simply evaluate $LL^-_{test} = LL^-(\mathbf{y}_{N_{train}:N_{train}+N_{test}}, \pmb{\theta})$. $N_{test}$ is the number of signal samples used and usually chosen as $N_{test} = N_{train}$.

If $LL^-_{test}$ evaluates to a similar value as $LL^-_{train}$, this is a good indication that a set of parameters was found that is general over the whole signal and a good optimum was reached.
A similar value is also a good indication that the choice of $N_{train}$ is reasonable. It means that the optimized parameters $\pmb{\theta}$ are generalized and represent a stationary equilibrium of the signal.

\subsection{Application to laser phase noise characterization}
\label{sec:application_to_laser}
To finalize the methodology section, we will step through a sample case of laser phase noise characterization in detail. The setup in \cref{fig:setup} will be simulated for the case where the LUT and LO both are single-frequency lasers. Subsequently, we will apply the conventional and EKF-based phase noise estimation methods and compare their accuracy.

\begin{figure}[ht]
    \centering
    \includegraphics[width=\linewidth]{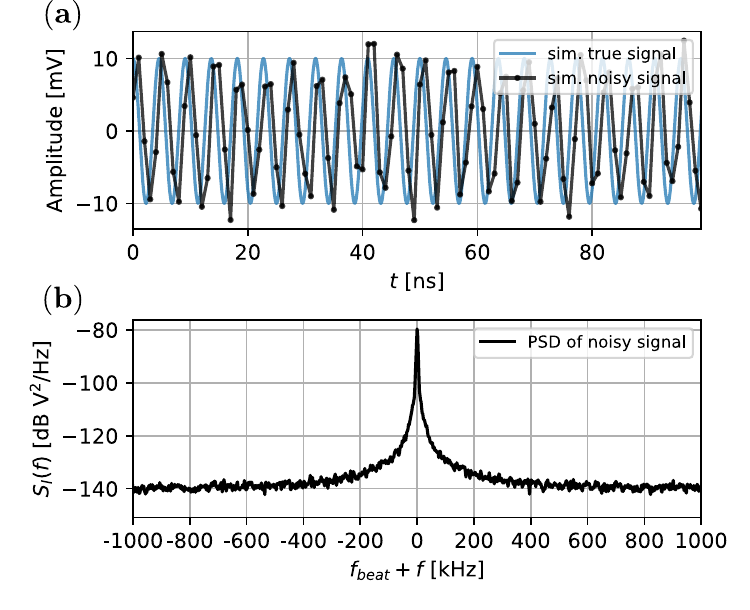}
    \caption{Simulated heterodyne laser beat note signal according to Eq. \eqref{eq:laser_beat_measured_signal}.
    (a) Time domain of the signal. The true signal without noise is an ideal sin function, the noisy signal exhibits strong perturbations.
    (b) PSD of the noisy signal. A strong peak at the beat frequency $\omega_{beat}$ with wide tails caused by phase noise is observed. The white measurement noise causes a flat noise floor at \SI{-140}{\decibel\volt^2\per\hertz}.
    }
    \label{fig:laser_beat_sim_signal}
\end{figure}
\subsubsection{Heterodyne laser beat note signal simulation}
To simulate the heterodyne setup, we use the generalized model of the electric field given in Eq. \eqref{eq:E_field_general}. Both the LUT and LO are single-frequency lasers (as opposed to combs) in this case, the general expression therefore collapses to 
\begin{align}
    E_{LUT}(t) &= \sqrt{P^{LUT}} e^{i(\omega^{LUT} t + \phi^{LUT}(t))} \\
    E_{LO}(t) &= \sqrt{P^{LO}} e^{i(\omega^{LO} t + \phi^{LO}(t))} \label{eq:E_field_laser_LO}
\end{align}
where amplitude noise is neglected. Each field only contains a single optical frequency that is perturbed by phase noise.
After optical coupling, balanced detection, and digitization by the ADC, the signal has the form
\begin{equation}
    \begin{split}
    y(t_k) &= 2\eta \sqrt{P^{LUT} P^{LO}}  \\
    &\qquad \times \sin{\left[(\omega^{LUT} - \omega^{LO})t_k + \phi^{LUT}(t_k) - \phi^{LO}(t_k)\right]} \\
    &\qquad + \xi(t_k) \\
    &\approx A \sin{\left[\omega_{beat} t_k + \phi^{LUT}(t_k)\right]} + \xi(t_k) \label{eq:laser_beat_measured_signal}
\end{split}
\end{equation}
in analogy to Eq. \eqref{eq:BPD_no_rin}.
For this simulation study, we assume the reference laser phase noise $\phi^{LO}(t_k)$ to be negligible, such that $\phi^{total}(t_k) \approx \phi^{LUT}(t_k)$.
As the simulation parameters we choose the beat note amplitude $A = \SI{10}{\milli\volt}$, the beat frequency $\omega_{beat} = \SI{220}{\mega\hertz}$ and a sampling frequency of $f_{sampling} = \SI{1}{\giga\hertz}$.
Both the measurement noise $\xi(t_k)$ and the phase noise $\phi^{LUT}(t_k)$ are random variables. The measurement noise samples $\xi_k$ have zero auto-correlation in time and are independently drawn from a zero mean normal distribution with variance $\sigma_R^2 = 5\cdot10^{-6}$. This corresponds to a measurement noise PSD level of \SI{-140}{\decibel\volt^2\per\hertz}, which is a typical experimental value for state-of-the-art measurement equipment.

It should be noted that in the heterodyne beat note case where the signal as expressed in \eqref{eq:laser_beat_measured_signal} is just a single sinusoidal, it is also possible to demodulate the signal experimentally. This would result in a measured signal $y(t_k)$ which is linear in $\phi^{LUT}(t_k)$ and would therefore not require the full EKF formalism. Instead, the (linear) Kalman filter could be used, simplifying the DSP. However, in the general multi-heterodyne case, this is no longer possible, which is why we introduce the non-linear EKF formalism here.

Laser phase noise typically has a non-zero time auto-correlation function. Most rigorously it can be generated by solving stochastic differential laser rate equations that model the LUT \cite{kantner_accurate_2023}.
For simplicity, however, we consider a phenomenological model for the frequency noise PSD of the phase noise of the laser. Using
\begin{align}
    S^{LUT}_{f}(f) = \Xi_{flicker} f^{-1} + \Xi_{Lorentz},\label{eq:laser_phase_noise_model}
\end{align}
the frequency noise PSD of the phase noise consists of a low-frequency flicker noise contribution $\Xi_{flicker} f^{-1}$ and a Lorentzian contribution $\Xi_{Lorentz}$ \cite{belhaq_bayesian_2024}. The frequency-independent term $\Xi_{Lorentz}$ models the theoretical phase noise contribution of an ideal quantum noise limited laser cavity \cite{schawlow_infrared_1958}. This term causes the line shape to have a Lorentzian profile. In addition, it can be used to define an intrinsic linewidth of a laser as $\Delta\nu = \pi \Xi_{Lorentz}$.
In real-world systems, low-frequency perturbations typically cause additional laser phase noise, which is reflected in the flicker noise term.

To generate a time series realization of $\phi^{LUT}(t_k)$ that follows the model in Eq. \eqref{eq:laser_phase_noise_model}, we use the method given in \cite{owens_algorithm_1978}.
In this study we choose $\Xi_{flicker}=10^6\;\si{\hertz\squared}$ and $\Xi_{Lorentz}=100 / \pi\; \si{\hertz}$ to generate $\phi^{LUT}(t_k)$.
Substituting into Eq. \eqref{eq:laser_beat_measured_signal} produces the simulated signal $y(t_k)$, which is displayed in Fig. \ref{fig:laser_beat_sim_signal}, together with its PSD.

\begin{figure}[t]
    \centering
    \includegraphics[width=\linewidth]{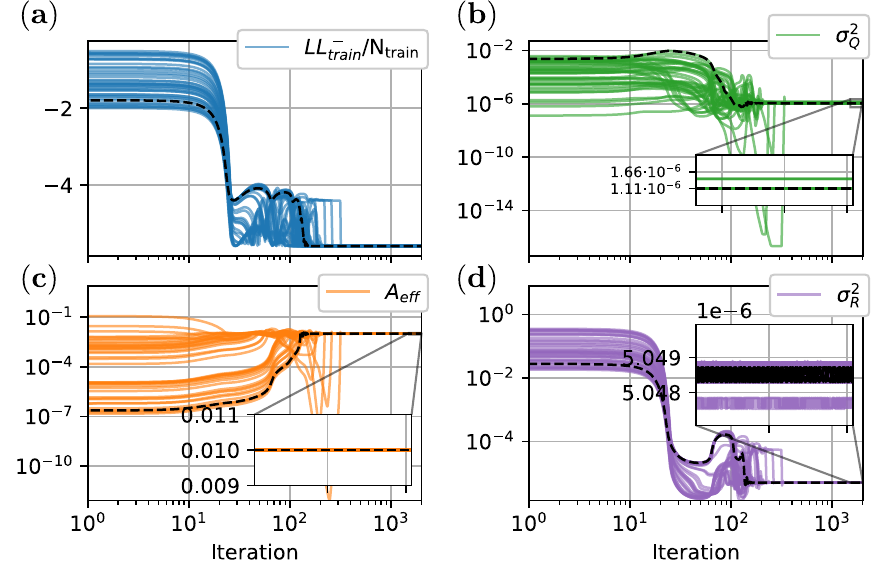}
    \caption{Convergence behaviour of selected parameters $\pmb{\theta}$ in the optimization process. 100 independent optimization runs with $L_{train}=\si{30000}$ were started with randomly drawn initial parameters $\pmb{\theta}_0$, of which the 40 best-performing ones are shown here. The performance is based on the final value of the cost function $LL^-$, which is shown in (a).
    (b,d) show the convergence of the phase and measurement noise variance, respectively.
    In (c) the convergence of the beat note amplitude is shown.
    While all runs converge on a single value of the amplitude, it appears that two distinct sets of final noise variances are found, depending on the initial parameters. This indicates that even for the case of a simple laser beat note measurement, $LL^-$ has local optima that can complicate convergence. The best optimization run is indicated by the black dashed line.}
    \label{fig:laser_beat_convergence}
\end{figure}

\subsubsection{EKF-based phase noise estimation}
Next, we apply the EKF-based DSP algorithm detailed in \cref{sec:EKF_method} to extract the phase noise given the simulated signal $y(t_k)$.
The first step is the definition of a state-space model. We first define a measurement function $\mathbf{h}$ that describes $y(t_k)$. As the analytical expression in Eq. \eqref{eq:laser_beat_measured_signal} was used to simulate $y(t_k)$, we can use this expression as the measurement function.
Reformatting Eq. \eqref{eq:laser_beat_measured_signal} to comply with the notation used in \cref{sec:EKF_method} yields
\begin{equation}
\begin{array}{llcr}
    y_{k} &= \mathbf{h}(x_k, \pmb{\theta}) + r_k &\qquad& \\
        &= A \sin{\left[\omega_{beat} t_k + x_k\right]} + r_k, & r_k \sim \mathcal{N}(0, \sigma_{R}).
    \label{eq:laser_beat_sim_model_h}
\end{array}
\end{equation}

Here, $x_k$ is the hidden state of the EKF and corresponds to the scalar-valued phase noise $\phi^{LUT}(t_k)$. $r_k$ represents the scalar-valued measurement noise $\xi(t_k)$ with variance $\sigma_{R}$.

\begin{figure}[t]
    \centering
    \includegraphics[width=\linewidth]{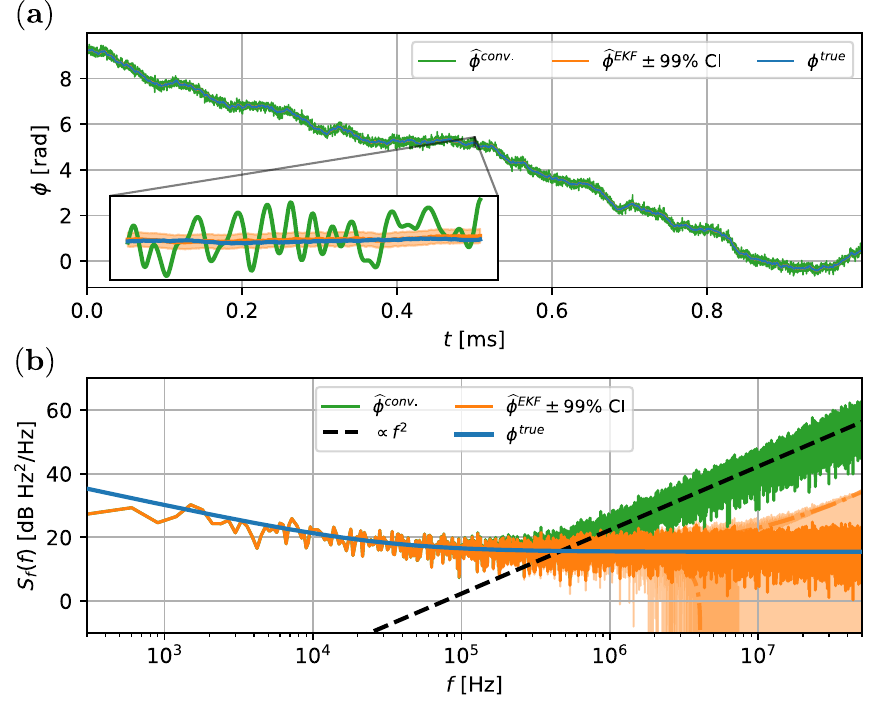}
    \caption{Comparison of the phase noise estimation methods in a simulated heterodyne laser beat note setup.
    (a) Time evolution of the phase noise as estimated by the conventional method and the EKF compared to the ground truth. The EKF's uncertainty is quantified with 99\% confidence intervals (CIs).
    (b) Frequency noise PSDs of the phase noise estimates. The true PSD follows the model given in Eq. \eqref{eq:laser_phase_noise_model}.}
    \label{fig:laser-beat-sim-phase-noise-PSD}
\end{figure}

The other half of the state-space model that needs to be defined is the state-transition function $\mathbf{f}$.
It models the time evolution of the phase noise and should ideally be as close as possible to the true time evolution.
In this simulation, the true phase noise model is given in Eq. \eqref{eq:laser_phase_noise_model}. We will, however, assume that the model is unknown to us and use the random walk approximation introduced in \cref{sec:state_space_model}.
Therefore, the state-transition function is given by 
\begin{equation}
\begin{array}{llcr}
    x_{k+1} &= \mathbf{f}(x_k, \pmb{\theta}) + q_k &\qquad& \\
        &= x_{k} + q_k, \qquad\qquad\qquad& q_k \sim \mathcal{N}(0, \sigma_{Q})
    \label{eq:laser_beat_sim_model_f}
\end{array}
\end{equation}
where the quantities in Eq. \eqref{eq:random_walk_model} have been replaced by their scalar-valued equivalents. 

With the state space model given in Eqs. \eqref{eq:laser_beat_sim_model_h} and \eqref{eq:laser_beat_sim_model_f}, its static parameters need to be optimized.
The parameters to be optimized in this model are $\pmb{\theta} = \{A, \delta_{\omega}, x_0, \sigma_{P_0}, \sigma_Q, \sigma_R\}$ where the initial phase $x_0$ and its variance $\sigma_{P_0}$ as well as the noise variances of the state space model were included.
Instead of optimizing the beat note frequency $\omega_{beat}$ directly, an initial estimate $\widetilde{\omega}$ is obtained by running a peak finding routine on the signal PSD displayed in Fig. \ref{fig:laser_beat_sim_signal}. We then define $\omega_{beat}=\widetilde{\omega}+\delta\omega$, where $\widetilde{\omega}$ is fixed, and $\delta\omega$ allows for fine-tuning of the frequency.
To ensure a good convergence behavior in the optimization and to constrain certain parameters, we define the parameter-wise optimization transformation function
\begin{align}
    \Lambda = 
    \begin{cases}
        \Lambda_A: A \rightarrow 10^{A}\\
        \Lambda_{\delta\omega}: \delta\omega \rightarrow \delta\omega_{max}\cdot \mathrm{tanh}(\delta\omega) \\
        \Lambda_{x_0}: x_0 \rightarrow x_0\\ 
        \Lambda_{\sigma_{Q}, \sigma_{R}, \sigma_{P_0}}: \sigma \rightarrow 10^\sigma
    \end{cases} \label{eq:optimization_transform_func_laser_beat}
\end{align}
which was introduced in \cref{sec:model_optimization}.
Here the exponential function with base ten is used to constrain the amplitude and the variances to non-negative values. $\delta\omega$ is constrained to values between $-\delta\omega^{max}$ and $\delta\omega^{max}$ using the hyperbolic tangents function. We choose $\delta\omega^{max}=\SI{5}{\kilo\hertz}$ as the initial estimate $\widetilde{\omega}$ is quite accurate.

With the state model and the parameter transformation function defined, the optimization loop illustrated in Fig. \ref{fig:setup} is executed using the AdaBelief optimizer.
To minimize the chance of sub-optimal convergence by running into a local optimum, 100 sets of randomly drawn initial parameters $\pmb{\theta}_0$ are chosen. Each parameter set is independently optimized until the change in parameters per iteration is negligible.
The convergence behavior for a selected subset of parameters is illustrated in Fig. \ref{fig:laser_beat_convergence}.
The initial values are distributed over a wide range of multiple decades to demonstrate the optimization routine's robustness.
Some of the optimization runs for which the random initialization of $\pmb{\theta}_0$ is particularly poor do not converge.
Most runs, however, converge to one of two local optima, which can be seen in the two distinct final values for the variances in Fig. \ref{fig:laser_beat_convergence} (b,d). Ultimately, only the final parameters $\pmb{\theta}$ of the best-performing run are used.

Using the optimal parameters $\hat{\pmb{\theta}}^{MAP}$, we can apply the EKF to estimate the time evolution of the phase noise as the final step.
We receive a time series of estimates $x_k$ with a corresponding time series of variances $\sigma^2_{P,k}$. Together they form a time series of Gaussian distributions $\mathcal{N}(x_k, \sigma_{P,k}^2)$ of near-optimal estimations of the phase noise time evolution.
To put the estimation into perspective, we apply the conventional phase estimation algorithm described in section \ref{sec:hilbert-method} on the same simulated signal.
Both methods are compared to the true phase noise in Fig. \ref{fig:laser-beat-sim-phase-noise-PSD}. The time domain traces shown in (a) indicate that the long-term behavior of the phase is evaluated accurately by both methods. Zooming in, however, reveals that the EKF estimate is more accurate than the conventional method.
The frequency noise PSD picture in Fig. \ref{fig:laser-beat-sim-phase-noise-PSD} (b) shows that the low-frequency dynamics up to \SI{300}{\kilo\hertz} are accurately captured by both methods.
For higher offsets, the conventional method is limited by the measurement noise floor $\xi(t_k)$, which causes the estimate to have a $\propto f^2$ frequency dependency. The EKF estimate follows the ground truth below the noise floor, demonstrating that the measurement noise is effectively filtered out.
The limits of the estimation accuracy are quantified by the variances $\sigma^2_{P,k}$ of the estimates. Using second-order error propagation, the time domain uncertainty can be propagated to the PSD estimate, allowing us to draw a CI around the PSD. A derivation is presented in the appendix.

In summary, the presented method using the EKF produces a more accurate estimate than the conventional method. The introduction of adaptive optimization algorithms and the use of AD make the method very versatile, as state space models of arbitrary form with arbitrarily many parameters can be efficiently optimized. In the following section, we will demonstrate how this framework can be used to estimate phase noise in frequency combs.

\section{Down-converted combs}
\label{sec:03_downconverted_combs}

\begin{figure}
    \centering
    \includegraphics[width=0.90\linewidth]{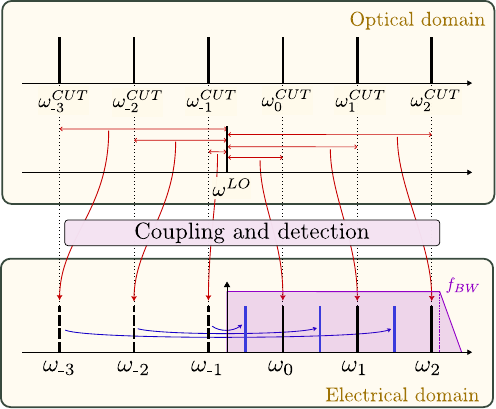}
    \caption{Illustration of the down-conversion process expressed in Eq. \eqref{eq:downconverted_comb_signal}, in which the optical comb signal is converted into the detected electronic signal. In the upper panel, the comb lines of the CUT and the LO are schematically displayed. As these are optical frequencies, they are offset hundreds of \si{\tera\hertz} from zero. Through the mixing and detection process, the differences of the optical frequencies are detected electrically. Note that negative difference frequencies are projected onto their corresponding positive frequencies through aliasing (blue arrows), as the signal we detect is real-valued. This is indicated by the purple overlay that illustrates the detected bandwidth. It is crucial to identify beat notes that have been aliased as their negative frequency needs to be part of the state space model for the EKF.}
    \label{fig:downconverted_comb_notes_origin}
\end{figure}

Two main options are available to characterize frequency combs in the setup described in Fig. \ref{fig:setup}. In this section, we will focus on the case where the CUT is a comb and the LO is a single-frequency laser. We will refer to this scenario as the "down-converted comb" setup in this chapter. The general expression for the measured signal from Eq. \eqref{eq:BPD_no_rin} reduces into
\begin{align}
    \begin{split}
        y(t_k) &= 2\eta \sum_{l=0}^\infty \sqrt{P^{CUT}_l P^{LO}}  \\
        &\quad \times \text{sin}[(\omega^{CUT}_{ceo} + l\omega^{CUT}_{rep} - \omega^{LO})t_k\\
        &\qquad  + \phi^{CUT}_l(t_k) - \phi^{LO}(t_k)] + \xi(t_k)\\
         &= \sum_{l=0}^\infty A_l \text{sin}[\omega_{l} t_k + \phi^{total}_l(t_k)] + \xi(t_k) 
    \end{split}
    \label{eq:downconverted_comb_signal}
\end{align}
where all quantities are the same as in Eq. \eqref{eq:BPD_no_rin}. As the LO is a single-frequency laser, one of the sums of the original expression is dropped. For compact notation we define $A_l=2\eta \sqrt{P^{CUT}_l P^{RS}}$, $\omega_l=\omega^{CUT}_{ceo} + l\omega^{CUT}_{rep} - \omega^{LO}$ and $\phi^{total}_l(t_k) = \phi^{CUT}_l(t_k) - \phi^{LO}(t_k)$. This general expression describes a multi-heterodyne measurement of any down-converted comb.

The beat note generation process described in Eq. \eqref{eq:downconverted_comb_signal} is illustrated in Fig. \ref{fig:downconverted_comb_notes_origin}.
The resulting signal is once again a comb with the same spacing as the original line spacing. To simultaneously detect as many lines as possible, we want to maximize the number of down-converted lines within the detection bandwidth. This is achieved by selecting a LO with a suitable frequency $\omega^{LO}$ which is close to the frequency of the highest power comb lines of the CUT. All lines for which $|\omega_l| < 2\pi f_{BW}$ can then be detected.
The maximum number of detectable lines within the detection bandwidth $f_{BW}$ is $N_{detected} \leq 2\lfloor 2\pi f_{BW} / \omega_{rep} \rfloor + 1$.
While we can always detect at least one comb line, this relation clearly illustrates that this setup is only suitable for combs that have a repetition frequency significantly below the available bandwidth. Only then can we detect multiple lines at the same time and benefit from the reduced experimental complexity of the setup to infer information about the comb phase noise from only a single measurement.

This condition disqualifies this setup for types of combs that have a repetition frequency $\omega_{rep}$ in the tens of \si{\giga\hertz} and above, as even state-of-the-art photodiodes are limited to around 100 \si{\giga\hertz} of bandwidth \cite{eng_state---art_2015}. Even with these extreme bandwidths, only a few comb lines would be detectable. Therefore, combs with large spacing can effectively only be characterized in dual-comb setups, which are discussed in \cref{sec:04_dual_comb}.

Further, this setup only has limited usability for combs with many lines on the order of thousands, which includes femtosecond laser-based combs or generally octave-spanning combs. Their repetition rate is typically on the order of \si{\giga\hertz} or less and beating with a single laser will produce many detectable beat notes. However, a single measurement will only ever cover a limited range of lines of the comb under test. For a full characterization of all lines, multiple multi-heterodyne measurements with a wavelength-tuneable LO would be required to access the phase noise of all lines. While not infeasible, a dual-comb setup with a suitable reference comb is preferable in this scenario.

For combs with $\omega_{rep}$ up to \SI{10}{\giga\hertz} and few comb lines, however, utilizing this single-frequency laser setup is more advantageous compared to employing a dual-comb configuration. With a single-frequency laser acting as the local oscillator (LO), it is still possible to detect numerous lines while the experimental setup is simplified significantly.
Furthermore, assessing the contribution of the LO towards the total phase noise per beat note $\phi^{total}_l(t_k)$ is more straightforward. As derived in Eq. \eqref{eq:downconverted_comb_signal} the LO phase noise $\phi^{LO}(t_k)$ is an additive contribution independent of the comb line index $l$. It, therefore, only adds a common phase noise offset that does not add any index-dependent dynamics.
This distinction is generally not as easy in a dual-comb setup, as the LO itself is a comb and its phase noise is therefore dependent on the line index.

To show how the EKF-based method can be used to extract the phase noise terms $\phi^{total}_l(t_k)$ in this scenario, we will first demonstrate its application on a simulated signal. An electro-optic (EO) comb is chosen as the CUT, as it is well-studied in literature and has a known phase noise model.
After showcasing the method on simulated data, the method's application is also demonstrated on experimental data.

\begin{figure}
    \centering
    \includegraphics[width=0.95\linewidth]{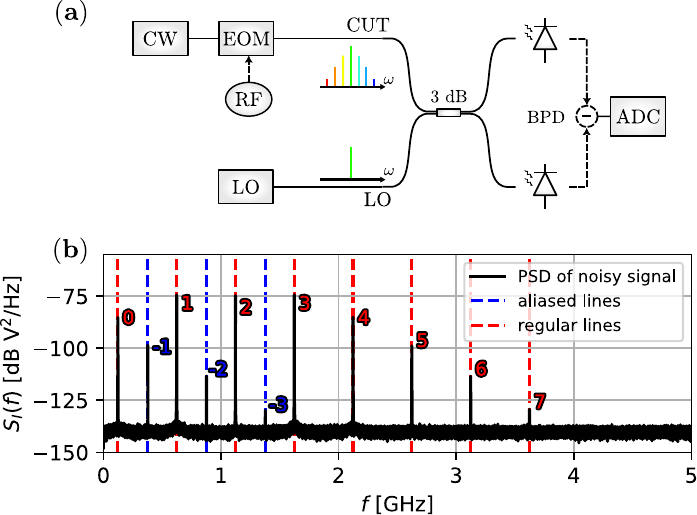}
    \caption{(a) Setup for the generation of an EO comb and subsequent detection of a multi-heterodyne beat note signal.
    CW: continuous wave laser, EOM: electro-optic modulator, RF: radio-frequency generator.
    (b) PSD of a simulated multi-heterodyne signal of an EO comb. The colored numbers correspond to the comb line index $l$, where negative indices indicate aliased lines.
    }
    \label{fig:downconverted_comb_sim_signal}
\end{figure}

\subsection{Simulating an EO comb}
\label{sec:EO_comb_sim}
The setup that is used to simulate a down-converted EO comb signal is displayed in \cref{fig:downconverted_comb_sim_signal} (a). The EO comb is generated by applying a sinusoidal phase modulation to a single-frequency seed laser (CW). Both the seed laser as well as the RF source that drives the electro-optica modulator are considered to exhibit phase noise.
The electric field of this EO comb has the form \cite{parriaux_electro-optic_2020}
\begin{align}
    E_{CUT}(t) &= \sqrt{P^{CW}} e^{i(\omega^{CW} t + \phi^{CW}(t))} e^{i K V(t)} \\
    V(t) &= V_0 \sin{\left[(\omega^{RF} t + \phi^{RF}(t)) \right]}. \label{eq:RF_voltage}
\end{align}
Here, $V(t)$ is the sinusoidal voltage that is generated by the RF source and applied to the EOM, which has a peak voltage $V_0$ and the frequency $\omega^{RF}$. The EOM acts as a pure phase modulator.
$K$ is a material-dependent proportionality constant of the EOM which determines the phase change per applied voltage.
The LO is simply a single-frequency laser, and its electric field $E_{LO}(t)$ is given by Eq. \eqref{eq:E_field_laser_LO}.

As detailed in \cref{sec:setup}, the electric fields $E_{CUT}(t)$ and $E_{LO}(t)$ are propagated through the setup using Eqs. \eqref{eq:optical_coupling} and \eqref{eq:BPD_with_RIN}. After balanced coupling, detection in the BDP and digitization by the ADC, the final signal then has the form
\begin{align}
\begin{split}
    y(t_k) &= 2\eta \sqrt{P^{CW} P^{LO}} \text{sin}\left[(\omega^{CW} - \omega^{LO}) t_k + K V(t_k) \right. \\
    &\qquad \left.+ \phi^{CW}(t_k) - \phi^{LO}(t_k)\right] + \xi(t_k) \label{eq:BPD_downconverted_comb}
    \end{split}
\end{align}
In this simulation study we neglect amplitude noise, and we assume that the phase noise $\phi^{LO}(t_k)$ of the LO is insignificant compared to the phase noise of the seed laser $\phi^{CW}(t_k)$ for simplicity such that $\phi^{CW}(t_k) - \phi^{LO}(t_k) \approx \phi^{CW}(t_k)$. For the general case where the LO noise can not be neglected, we will keep $\phi^{LO}(t_k)$ as part of the model throughout this section.
For $\phi^{CW}(t_k)$, we use the same phase noise model as in the laser beat note study in \cref{sec:application_to_laser}.
We use the model described by Eq. \eqref{eq:laser_phase_noise_model}, in which the phase noise is composed of a random walk with added low-frequency flicker noise. In this study we choose $\Xi_{flicker}=10^6\;\si{\hertz\squared}$ and $\Xi_{Lorentz}=10^3/\pi\;\si{\hertz}$ as the parameters for the model.

The third phase noise term that appears in Eq. \eqref{eq:BPD_downconverted_comb} is hidden in $V(t_k)$, the output of the RF source. As Eq. \eqref{eq:RF_voltage} states, $\phi^{RF}(t_k)$ is the phase noise in the RF signal. RF source phase noise generally has different and more complicated characteristics compared to laser phase noise. Therefore, we will use a different phenomenological model to generate $\phi^{RF}(t_k)$.

Ultra-low phase noise RF sources often have multiple plateaus in their phase noise PSD caused by multiple independent locking mechanisms \cite{hajimiri_noise_2001}. A physically accurate simulation would heavily depend on implementation details and is out of the scope of this work.
To still roughly emulate this behavior and to intentionally deviate from the laser phase noise model, $\phi^{RF}(t_k)$ is simulated using 
\begin{align}
    S^{RF}_{f}(f)= \Xi_{Plateau,RF}\left(\sin{\left(\log_{10}(\alpha f)\right)}\right)^8 + \Xi_{Lorentz,RF}. \label{eq:RF_pn_model}
\end{align}
which has two free scale parameters $\Xi_{Lorentz,RF}$ and $\Xi_{Plateau,RF}$. The former models an underlying random walk of the phase noise. The latter models the scale of a sinusoidal deviation that is added to the random walk.
The factor $\alpha=\SI{1}{\hertz^{-1}}$ only ensures that the argument to the logarithm and sine are unitless.
For this numerical study, we choose the free parameters as $\Xi_{Lorentz,RF} = \SI{100}{\hertz}$ and $\Xi_{Plateau,RF} = 10^4\;\si{\hertz}$ with the corresponding unit, which creates a phase noise PSD model that has multiple plateaus.

Note that this model and its parameters are phenomenologically chosen to emulate the phase noise of a specific device, which we have used as a reference \footnote{see Agilent E8257C datasheet}.
The secondary purpose of this model is its explicit deviation from a simple phase noise random walk.
As discussed in \cref{sec:application_to_laser}, the EKF state-transition function we use to approximate the phase noise behavior does not consider any deviations from a random walk. An intentional mismatch between the true and approximated model can, therefore, give an indication of whether the random walk state-space model can still capture complicated phase noise dynamics.

To simulate time series realizations $\phi^{CW}(t_k)$ and $\phi^{RF}(t_k)$ based on the PSD models in Eqs. \eqref{eq:laser_phase_noise_model} and \eqref{eq:RF_pn_model}, we again use the algorithm given in \cite{owens_algorithm_1978}.
Eq. \eqref{eq:BPD_downconverted_comb} can now be used to directly simulate a time domain signal of a down-converted EO comb.
The parameters chosen for this numerical study are 
$2\eta \sqrt{P^{CW} P^{LO}}=\SI{100}{\milli\volt}$, $K V_0=\num{1,6}$, $\omega^{CW}=2\pi\cdot\SI{1.1}{\giga\hertz}$ and $\omega^{RF}=2\pi\cdot\SI{500}{\mega\hertz}$. Here, we have combined some individual parameters into effective combined quantities, as only total values, as opposed to the individual contributions, matter.
The spectrum of the resulting signal $y(t_k)$ is shown in \cref{fig:downconverted_comb_sim_signal} (b). The resulting comb is centered around the line with index $l_{center}=2$ and features $N_{alias} = 3$ and $N_{reg.} = 8$ aliased and non-aliased comb lines, respectively.

\subsection{Latent phase noise state-space model}
To use the EKF-based phase noise estimation method, a state-space model needs to be defined.
The measurement equation of the model, which describes the detected signal, is based on the general analytical expression in Eq. \eqref{eq:BPD_downconverted_comb}.

The simplest approach to tracking the comb's phase noise is to track the phase noise terms of each beat note individually.
In analogy to the laser beat note case, the hidden variables of the state space model are the phase noise terms $\phi_l(t_k)$ of the individual beat notes. They are approximated by a multi-dimensional random walk.
The state-space model takes the form
\begin{align}
\begin{split}
    \mathbf{f}_{indiv.}(\mathbf{x}_k, \pmb{\theta}) &= \mathbf{x}_k + \mathbf{q}_k \\
    \mathbf{q}_k &\sim \mathcal{N}(0, \mathbf{Q}_N)
\end{split}  \label{eq:model_comb_transition}\\
\begin{split}
    \mathbf{h}_{indiv.}(\mathbf{x}_k, \pmb{\theta}) &= \sum_{l=-N_{alias}}^{N_{reg.}-1} A_{l} \text{sin}\left[ \left( \omega_{0} + l\omega_{rep} \right) t_k + \mathbf{x}_k \right] + r_k \\
    r_k &\sim \mathcal{N}(0, \sigma_R)
    \end{split}  \label{eq:model_comb_measurement}
\end{align}
where we identify the $N = N_{alias} + N_{reg.}$ dimensional vector of hidden variables $\mathbf{x}_k$ as the beat note phase noise terms $\pmb{\phi}(t_k)$. $N_{alias}$ and $N_{reg.}$ are the number of aliased and non-aliased comb lines with significant amplitude in the recorded spectrum, respectively.
$\omega_{0}$ is chosen as the lowest of the beat note frequencies that is not aliased (as indicated in Fig. \ref{fig:downconverted_comb_notes_origin}).
The static parameters are $\pmb{\theta}_{indiv.} = \{ \mathbf{A}, \omega_{0}, \omega_{rep}, \mathbf{x}_0, \mathbf{P}_0, \mathbf{Q}_N, \sigma_R\}$. This model has $|\pmb{\theta}_{individual}| = \nicefrac{1}{2} \left( N^2 + 7N \right) + 3$ free parameters that need to be optimized.

Using this state-space model for EKF-based phase noise tracking has been shown to produce accurate estimations that outperform the conventional method in an EO comb \cite{brajato_bayesian_2020}.
The approach does, however, have multiple drawbacks that have to be addressed.

In this individual line tracking approach, the correlations between the phase noise terms $\phi^{total}_l(t_k)$ are found through the optimization of $\mathbf{Q}_N$. This matrix represents the differential phase noise covariance, from which the differential correlations can be calculated.
No prior information about the phase noise in the system is used in this model.

The state vector $\mathbf{x}_k$ has the same dimension as the number of detected comb lines $N$. This causes the associated covariance matrices $\mathbf{P}_k$ as well as the process noise covariance matrix $\mathbf{Q}_N$ of the state vector to be a matrix of size $\left( N \times N\right)$. The EKF filtering equations \eqref{eq:EKF_equations} include matrix operations that have an undesirable computational complexity scaling of $\mathcal{O}(N^3)$. Applying the EKF on a large number of comb lines $N$ can make this approach computationally infeasible.
Additionally, the covariance matrix $\mathbf{Q}_N$ needs to be optimized, which requires the optimization of $\mathcal{O}(N^2)$ free parameters.

To solve the complexity scaling problem, we can exploit the fact that the individual comb line phase noise terms $\phi^{total}_l$ is a combination of few underlying phase noise sources.
The expression for the signal in Eq. \eqref{eq:BPD_downconverted_comb} shows that only three phase noise sources contribute. The $N$ individual beat note phase noise terms $\phi^{total}_l(t_k)$ are a combination of the three sources, which means that they are correlated.

For the EO comb, these correlations are well studied \cite{ishizawa_phase-noise_2013, deakin_phase_2021}. The phase noise of each beat note is
\begin{align}
    \phi^{total}_l(t_k) &= \underbrace{\phi^{CW}(t_k) - \phi^{LO}(t_k)}_{\phi^{comm}(t_k)} + ( l - l_{center})\phi^{RF}(t_k) \label{eq:EO_comb_phase_noise}
\end{align}

In matrix notation, it becomes
\begin{align}
    \pmb{\phi}^{total}(t_k) &= \mathbf{C}_{true} \begin{bmatrix} \phi^{comm}(t_k)\\ \phi^{RF}(t_k)\end{bmatrix} \label{eq:EO_comb_latent_space} \\
    \mathbf{C}_{true} &= \begin{bmatrix}
... & 1 & 1 & 1 & 1 & 1 &...\\
... & -2& -1 & 0 & 1 & 2 &...
\end{bmatrix}^T. \label{eq:C_true}
\end{align}
The second column of $\mathbf{C}_{true}$ is zero at the index of the central comb line $l_{center}$.
We have introduced a mixing matrix $\mathbf{C}$ which describes the linear transformation that connects the true phase noise sources with the phase noise per beat note. Independent of the number of beat notes $N$, each $\phi^{total}_l(t_k)$ is described by only two independent contributions: A common mode noise term $\phi^{comm}(t_k) = \phi^{CW}(t_k) - \phi^{LO}(t_k)$ that is added to all beat notes independent of $l$, and a differential mode noise term $\phi^{RF}(t_k)$ which has a linear dependence on $l$.

The structure of Eq. \eqref{eq:EO_comb_latent_space} can be used to design a more efficient state-space model.
Instead of tracking the phase noise of each beat note, we can track the phase noise sources directly. Using Eq. \eqref{eq:EO_comb_latent_space}, $\mathbf{\phi}^{total}(t_k)$ can be constructed if $\mathbf{C}_{true}$ is known.
In more formal terms, we introduce a linear latent space into the state-space model, which is a lower-dimensional representation of the comb line phase noises $\mathbf{\phi}^{total}(t_k)$. In general, this can be expressed as 
\begin{align}
    \pmb{\phi}(t_k) = \mathbf{C} \mathbf{\psi}(t_k) \label{eq:latent_space_eq}
\end{align}
where $\mathbf{\psi}(t_k)$ is a $L < N$ dimensional vector of latent phase noise sources. 

Including this latent space into the measurement function of the state-space model given in Eq. \eqref{eq:model_comb_measurement} yields an improved model:
\begin{align}
\begin{split}
    \mathbf{f}_{lat.}(\mathbf{x}_k, \pmb{\theta}) &= \mathbf{x}_k + \mathbf{q}_k \\
    \mathbf{q}_k &\sim \mathcal{N}(0, \mathbf{Q}_L)
\end{split}  \label{eq:latent_model_comb_transition}\\
\begin{split}
    \mathbf{h}_{lat.}(\mathbf{x}_k, \pmb{\theta}) &= \sum_{l=-N_{alias}}^{N_{reg.}-1} A_{l} \text{sin}\left[ \left( \omega_{0} + l\omega_{rep} \right) t_k + \mathbf{C}\mathbf{x}_k + \phi^0_l \right] + r_k \\
    r_k &\sim \mathcal{N}(0, \sigma_R)
    \end{split}  \label{eq:latent_model_comb_measurement}
\end{align}
The hidden state $\mathbf{x}_k$ is the $L$ dimensional vector of latent phase noise sources $\mathbf{\psi}(t_k)$.
$\phi^0_l$ is a global initial phase of the \textit{l}-th line. While not strictly required as the initial values $\mathbf{x}_0$ also can contribute to an initial phase, the inclusion of $\phi^0_l$ improves optimization convergence.

\begin{figure*}
    \centering
    \includegraphics[width=\linewidth]{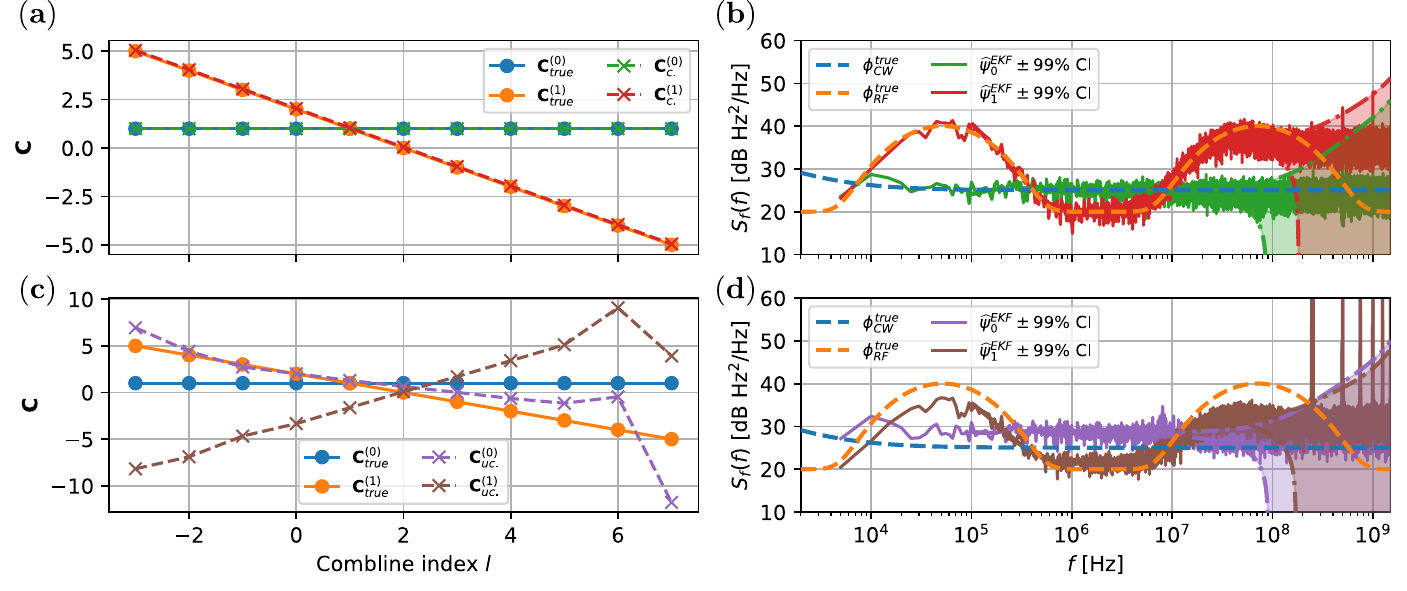}
    \caption{
    Application of the EKF-based phase noise estimation method on a simulated EO comb.
    (a) Column vectors (indicated by the superscripts $^{(0)}$ and $^{(1)}$) of the true and optimized mixing matrix $\mathbf{C}_{c.}$ which describe the latent phase noise space of the EO comb. During the optimization, $\mathbf{C}_{c.}$ is constrained to only exhibit common mode and differential mode phase noise.
    (b) Frequency noise PSDs $S_f(f)$ of the latent phase noise sources, based on the mixing matrix shown in (a).
    (c) Column vectors of the true and optimized mixing matrix $\mathbf{C}_{uc.}$ where the entries of the matrix are unconstrained during optimization.
    (d) Frequency noise PSDs $S_f(f)$ of the latent phase noise sources corresponding to the unconstrained mixing matrix shown in (c).
    All CIs shown are smoothed for visibility.
    }
    \label{fig:C_constrain_vs_unconstrain}
\end{figure*}

The static parameters of this model are $\pmb{\theta}_{lat.} = \{ \mathbf{A}, \mathbf{\phi}^0, \omega_0, \omega_{rep}, \mathbf{x}_0, \mathbf{P}_0, \mathbf{C}, \mathbf{Q}_L, \sigma_R\}$.
The dimension $L$ of the latent space is a hyperparameter of the model. This means that it can not be optimized directly, but instead needs to be chosen based on prior information.
In the EO comb case, Eq. \eqref{eq:EO_comb_latent_space} shows that a two dimensional latent space with $L=\num{2}$ can explain the phase noise of all comb lines. Further, the form of the mixing matrix $\mathbf{C}$ is given by Eq. \eqref{eq:C_true}. The only free parameter in this expression is the index of the central comb line, which shifts the zero-crossing point of the second column of the matrix. As $\mathbf{C}$ is part of $\pmb{\theta}_{lat.}$, the index offset can be optimized alongside the other parameters.
To perform the optimization, we define the optimization transformation function $\Lambda$ as introduced in \cref{sec:model_optimization}:
\begin{align}
    \Lambda(\pmb{\theta}) = \left\{
    \begin{array}{llrl}
        \Lambda_{A_l}&: &A_l &\rightarrow 10^{A_l}\\
        \Lambda_{\phi_l^0}&: &\phi_l^0 &\rightarrow \phi_l^0 \\
        \Lambda_{x_0}&: &x_0 &\rightarrow x_0\\ 
        \Lambda_{\delta\omega_0}&: &\delta\omega_0 &\rightarrow \delta\omega_{max}\cdot \mathrm{tanh}(\delta\omega_0) \\
        \Lambda_{\delta\omega_{rep}}&: &\delta\omega_{rep} &\rightarrow \delta\omega_{max}\cdot \mathrm{tanh}(\delta\omega_{rep}) \\
        \Lambda_{\mathbf{P}_{0,ij}}&: &\mathbf{P}_{0, ij} &\rightarrow \delta_{ij} 10^{\mathbf{P}_{0, ij}} \\
        \Lambda_{\mathbf{Q}_{L, ij}}&: &\mathbf{Q}_{L, ij} &\rightarrow \delta_{ij} 10^{\mathbf{Q}_{L, ij}} \\
        \Lambda_{\sigma_{R}}&: &\sigma &\rightarrow 10^\sigma \\
        \Lambda_{\mathbf{C}}&: &l_{center} &\rightarrow \begin{bmatrix}
1 & 0-l_{center} \\
1 & 1-l_{center} \\
1 & 2-l_{center} \\
\vdots & \vdots
\end{bmatrix}\\
    \end{array}\right. \label{eq:optimization_function_sim_eo_comb}
\end{align}
The function is very similar compared to the laser beat note case in Eq. \eqref{eq:optimization_transform_func_laser_beat}. A notable difference is that $\mathbf{P}_0$ and $\mathbf{Q}_L$ are matrices, as the hidden state dimension is larger than one. $\delta_{ij}$ is the Kronecker delta, which forces both $\mathbf{P}_0$ and $\mathbf{Q}_L$ to be diagonal matrices. This explicitly disallows correlations between the latent noise sources $\mathbf{\psi}(t_k)$.
While it is possible to allow non-zero off-diagonal elements, this complicates the optimization process, as $\mathbf{Q}_L$ has to be constrained to the space of positive semi-definite matrices as it is a covariance matrix. Only allowing diagonal matrices avoids this problem and additionally reduces the number of parameters to optimize.
This is equivalent to constraining the latent phase noise sources to be uncorrelated. For the EO comb this is a reasonable assumption as the true phase noise sources are also uncorrelated.
To optimize the comb frequencies, we again choose $\delta\omega^{max}=\SI{5}{\kilo\hertz}$.

The function $\Lambda_{\mathbf{C}}$ that transforms the mixing matrix is based on Eq. \eqref{eq:C_true}. Only a single free parameter, $l_{center}$, is required to define the matrix. The latent phase noise space is constrained to only have a common mode and a differential mode phase noise contribution. We will call this the constrained latent space model, indicated by the subscript $\mathbf{C}_{c.}$.

The constrained model has $|\pmb{\theta}_{c.}| = 2N + 3L + 4$ free parameters and $L$ tracked variables.
For large $N$ the linear scaling of the number of parameters causes it to have significantly fewer free parameters compared to the individual line approach.
Notably, the number of parameters no longer scales quadratically with the number of comb lines. Further, the EKF computational complexity reduces to $\mathcal{O}(L^3)$, which is no longer dominated by scaling with the number of comb lines.
This means that it becomes feasible to use this model to apply the EKF-based phase noise estimation method on combs with a large number of comb lines.

\paragraph*{Unconstrained latent spaces}
While the latent phase noise space of an EO comb can be explicitly described by Eq. \eqref{eq:C_true}, a similar expression might not be available for all types of combs.
In particular, evidence for latent phase noise sources beyond the common mode and differential mode has been reported for mode-locked lasers \cite{benkler_circumvention_2005, razumov_subspace_2023}.

In these scenarios, we can relax the requirement for prior information about the phase noise while still benefiting from the latent state-space model. Instead of using the transformation function $\Lambda_{\mathbf{C}}$ as defined in Eq. \eqref{eq:optimization_function_sim_eo_comb}, it can be replaced by the identity function
\begin{align}
    \Lambda_{\mathbf{C}}: \mathbf{C}_{lm} \rightarrow \mathbf{C}_{lm}. \label{eq:optimization_function_uc}
\end{align}
If this function is used instead, each entry of $\mathbf{C}$ becomes a free parameter of the state-space model. The latent space decomposition becomes fully data-driven as the optimizer learns the entries of $\mathbf{C}$. No assumptions about the comb line phase noise are made, except for the assumption that only a few latent sources contribute to each comb line.
We will call this variant the unconstrained model, indicated by the subscript $\mathbf{C}_{uc.}$.

In this form, the model has $|\pmb{\theta}_{uc.}| = NL + 2N + 3L + 3$ parameters, which still has a favorable scaling compared to the individual line tracking approach as $N$ becomes large. In addition, the complexity of applying the EKF is still $\mathcal{O}(L^3)$.

\subsection{Application on simulated EO combs}
To evaluate the latent space phase noise model, we apply it to the simulated EO comb discussed in \cref{sec:EO_comb_sim}.
We will compare two variants of the latent space model against the ground truth phase noise.

The constrained latent state-space model assumes that the comb line phase noise is explained by a common and a differential phase noise mode. This model is derived from the known behavior of EO comb phase noise as expressed in Eq. \eqref{eq:EO_comb_phase_noise}.
The state-space model is defined in \eqref{eq:latent_model_comb_transition} and \eqref{eq:latent_model_comb_measurement}. Together with the optimization transition function which is defined in Eq. \eqref{eq:optimization_function_sim_eo_comb}, the parameters $\pmb{\theta}_{c}$ are optimized, as detailed in \cref{sec:model_optimization}, and used to estimate the latent phase noise sources $\widehat{\boldsymbol{\psi}}^{c}(t_k)$.

For comparison, we also test the unconstrained state-space model variant on the same simulated signal. This variant uses the same state-space model. The difference lies in the optimization transformation function, where we use the replacement detailed in Eq. \eqref{eq:optimization_function_uc}. While the latent space is still kept two-dimensional, it is not constrained to common and differential noise modes. Using the optimization framework, we find the parameters $\pmb{\theta}_{uc.}$ and use them to estimate the latent phase noise $\widehat{\psi}^{uc.}_m(t_k)$.

\begin{figure*}[t]
    \centering
    \includegraphics[width=\linewidth]{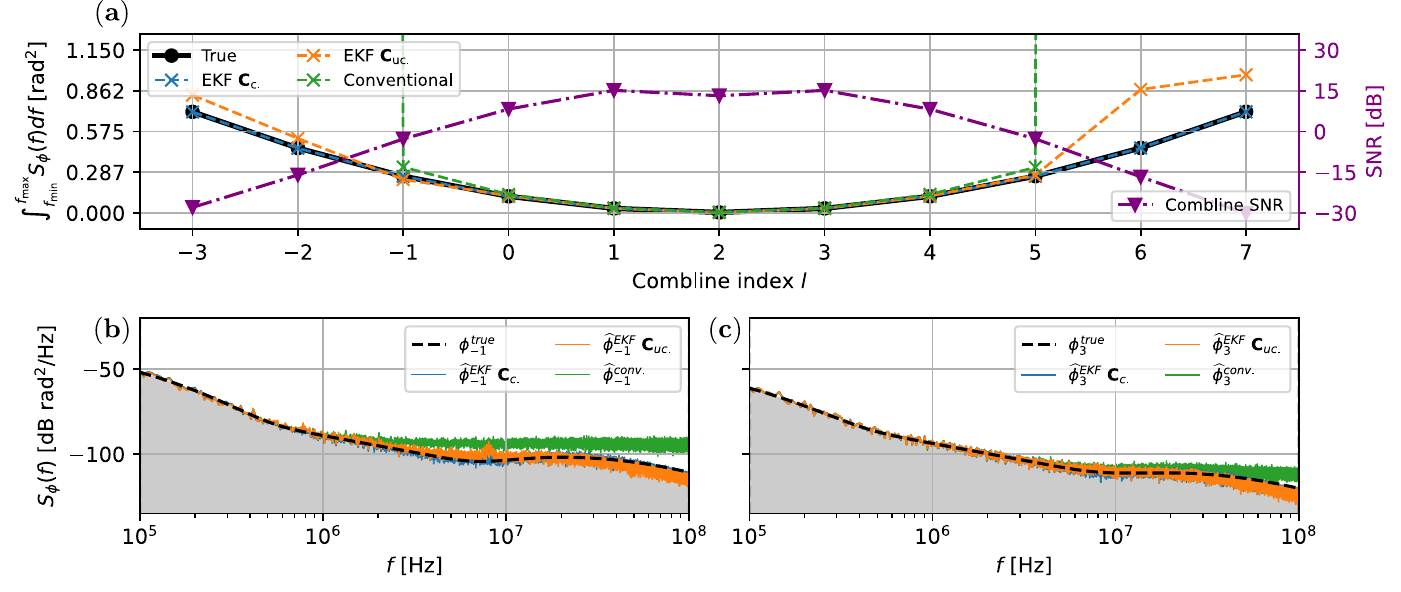}
    \caption{Accuracy comparison of the different phase noise estimation methods on a simulated EO comb.
    (a) Integrated phase noise PSD per comb line for each method. The estimates of the EKF using the constrained ($\mathbf{C}_{c.}$) and unconstrained ($\mathbf{C}_{uc.}$) latent space models, as well as the conventional method, are compared against the ground truth.
    Additionally, the SNR per comb line is shown with the purple, triangular markers which are measured on the right y-axis.
    (b) and (c) illustrate the phase noise PSDs of selected comb lines with indices $l=\num{-1}$ and $l=\num{3}$ respectively. The gray shaded area corresponds to the integrated PSD displayed in (a) for each line.
    }
    \label{fig:eo_comb_lines_variance}
\end{figure*}

\cref{fig:C_constrain_vs_unconstrain} displays the results of the EKF-based estimation.
In \cref{fig:C_constrain_vs_unconstrain} (a) the columns of the optimized mixing matrix $\mathbf{C}_{c.}$ are compared to the true matrix. The single free parameter of $\mathbf{C}_{c.}$ in Eq. \eqref{eq:optimization_function_sim_eo_comb} which identifies the index of the central comb lines was optimized. Its final value $l^{MAP}_{center} \approx \num{2}$ is very close to the true value. Therefore the columns of $\mathbf{C}_{c.}$ align perfectly with the true values. In (b), the frequency noise PSDs of the latent phase noise sources are shown. As $\mathbf{C}_{c.}$ matches $\mathbf{C}_{true}$, the latent noise sources map directly to the physical phase noises $\phi_{true}^{comm}(t_k) = \phi_{true}^{CW}(t_k) - \phi_{true}^{LO}(t_k)$ and $\phi_{true}^{RF}(t_k)$. At high offset frequencies $f>\SI{100}{\mega\hertz}$, the EKF estimates deviate from the true PSDs. The CI around the PSD, however, captures this estimation inaccuracy.

In \cref{fig:C_constrain_vs_unconstrain} (c) the optimized mixing matrix $\mathbf{C}_{uc.}$ of the unconstrained model is shown. The column vectors do not display a well-defined separation into common and differential noise modes. The latent phase noise sources will, therefore, not correspond to physical noise sources. This can be seen in (d), where their PSDs are shown. The shape of the PSDs are similar to the true ones, but they deviate considerably.
In addition, both latent phase noise PSDs display strong spikes at high frequencies which are not present in the true phase noise sources.
The origin of these high-frequency oscillations is unclear as they are not part of the state-space model. It is likely that due to the sub-optimal set of parameters $\pmb{\theta}_{uc.}$, the EKF produces erratic phase estimates to compensate for the accordingly inaccurate state-space model. This showcases the importance of finding good model parameters since the phase noise estimates cannot be trusted otherwise.

While it can be useful to isolate the physical noise sources in a comb, this is only possible if a prior model of the comb phase noise exists, as in Eq. \eqref{eq:EO_comb_phase_noise}. In all other cases, especially in the unconstrained mixing matrix case, the optimizer will find any $\mathbf{C}$ that can approximately decompose the comb line phase noise into latent noise sources. 

Here, the latent space is not interpreted as a physical quantity. However, the observable of interest are the phase noise per comb line $\phi^{total}_l(t_k)$. The latent space simply serves as a compressed representation. Using Eq. \eqref{eq:latent_space_eq}, $\phi^{total}_l(t_k)$ can easily be computed from the latent phase noise.

\subsubsection*{Estimation accuracy}
In \cref{fig:eo_comb_lines_variance} the estimation accuracy of both the constrained and unconstrained EKF-based approach and the conventional method are compared against the ground truth. 
The phase noise per comb line can also be estimated using the conventional method discussed in \cref{sec:hilbert-method}. The accuracy is estimated based on the estimated variance of the phase noise.

In (a) the integrated phase noise PSDs of each individual comb line for the different methods are shown. This measure is similar to the empirical phase noise time domain variance as it has the same unit, except that it considers a wide part of the spectrum instead. The phase noise PSD of each comb line $S_{\phi}^{total,l}(f)$ is integrated from $f_{min}=\SI{100}{\kilo\hertz}$ to $f_{max}=\SI{100}{\mega\hertz}$. The value of $f_{max}$ is chosen since \SI{100}{\mega\hertz} is the maximum offset frequency we can evaluate using the conventional method in this case. 

\begin{figure*}
    \centering
    \includegraphics[width=\linewidth]{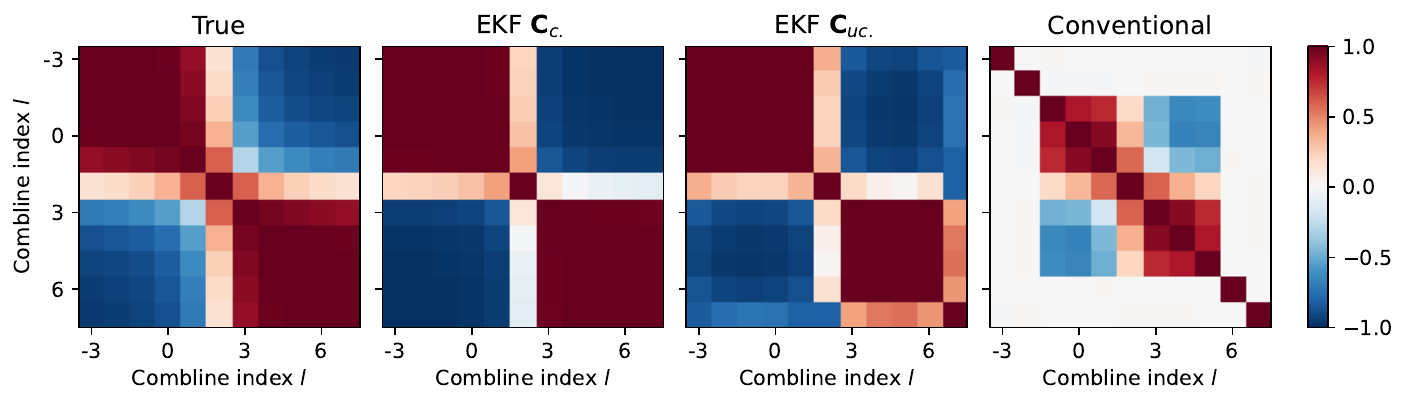}
    \caption{
    Estimated differential phase noise correlation matrices of a simulated EO comb. The matrices are estimated empirically from the time domain phase noise traces.
    }
    \label{fig:correlation_matrices}
\end{figure*}

The constrained EKF variant produces estimates that perfectly align with the true value. This is unsurprising as the latent space estimate shown in \cref{fig:C_constrain_vs_unconstrain} (a,b) is accurate as well.
However, the estimate also is accurate for the outermost lines, which have less than \SI{-20}{\decibel} SNR. 
Here we define the electrical SNR per beat note $l$ as
\begin{align}
    \text{SNR}_l = \frac{A_l^2 / 4}{\sigma_R^2 \Delta T_s f_{BW}} \label{eq:SNR}
\end{align}
where $A_l$ is its amplitude, $\sigma_R^2$ is the variance of the white measurement noise, $\Delta T_s$ is the sampling period and $f_{BW}$ is the detection bandwidth.
While this seems counterintuitive, this impressive measurement noise rejection can be explained by the restrictive state-space model.
Since $\mathbf{C}_{c.}$ is determined by only one parameter, there are not enough degrees of freedom in the model to allow for phase noise overestimation. Intuitively, the single free parameter is determined during the optimization based on the high SNR lines. The phase noise of the noisy low SNR lines is then extrapolated based on the constrained phase noise model, such that the low SNR of the lines does not impact the estimation process.

The unconstrained variant with $\mathbf{C}_{uc.}$ produces accurate estimates for the central comb lines which have high SNR. For the outer lines with $\text{SNR}<\SI{15}{\decibel}$, the estimated variance is larger than the true value, which indicates that the phase noise magnitude is overestimated.
The conventional method estimates are similar where the high SNR comb line phase noise is estimated accurately. For the outer lines, however, the method completely fails. The very low SNR indicates that the measurement noise dominates over the beat notes, a regime in which the conventional method can not produce a good estimate.

In \cref{fig:eo_comb_lines_variance} (b,c) two exemplary phase noise PSD of the comb lines with indices $l=-1$ and $l=3$ are shown. The former beat note has low SNR, which is, however, still enough such that the conventional method does not fail completely. Beyond $f=\SI{2}{\mega\hertz}$ its PSD estimate is dominated by measurement noise. Both EKF variants follow the true PSD, which is below the noise floor.
The beat note with index $l=3$ is one of the highest SNR notes. The conventional estimate is considerably closer to the true PSD, as the noise floor is lower than that of line $l=-1$.

\begin{figure*}
    \centering
    \includegraphics[width=\linewidth]{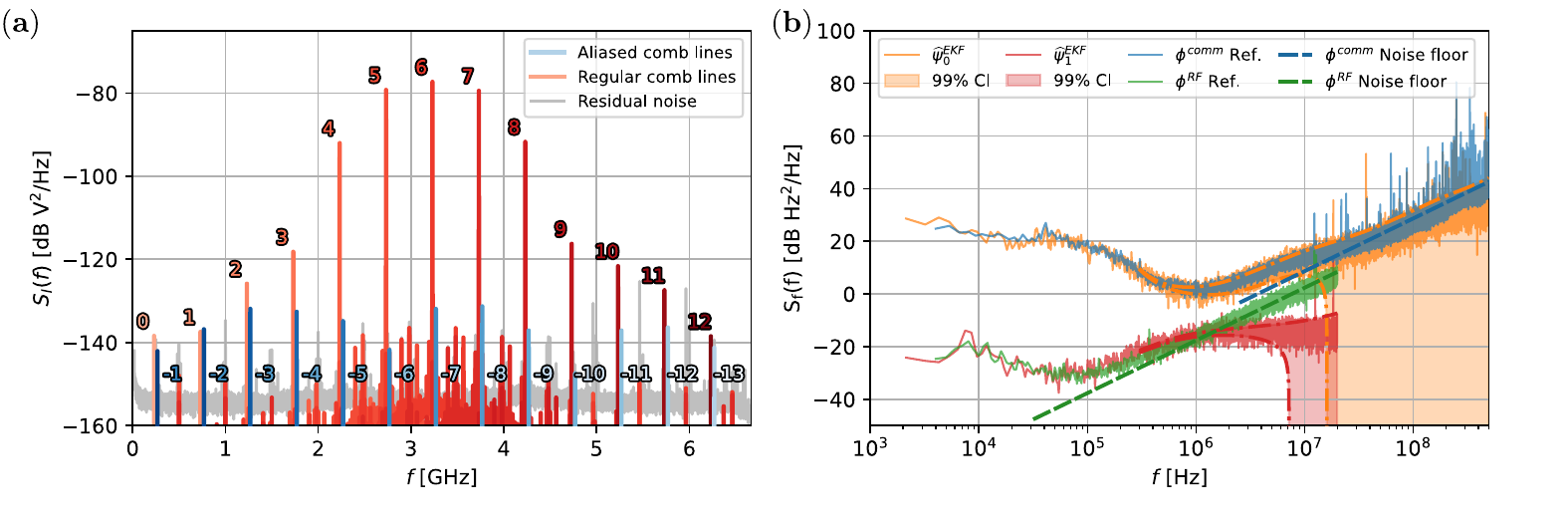}
    \caption{
    (a) PSD of an experimental mutli-heterodyne EO comb signal. The number next to each line is its index, where negative numbers indicate aliased lines.
    To showcase the EKF's ability to decompose the signal, the contribution from each line is isolated and displayed in a slightly different color. Individual comb lines can be isolated without the need for a bandpass filter. The signal, as detected in the experiment, is the sum of all lines and the residual noise.
    (b) Frequency noise PSDs of the latent phase noise sources, as estimated by the EKF. The PSDs are shown with a 99\% CI. Also shown are frequency noise PSDs of the common mode noise originating from the lasers and RF source noise used in the EO comb generation, extracted from independent reference measurements. The measurement noise floors of the reference measurements are indicated with dashed lines.
    }\label{fig:exp_eo_comb_signal_and_latent_psds}
\end{figure*}

\subsubsection*{Correlation matrices}
Finally, we investigate the correlations of the comb lines phase noise in the time domain.
First, the covariance matrices are estimated using the sample covariance matrix defined as
\begin{align}
    \widehat{\mathbf{Q}}_N &= \frac{1}{1-T} \sum_{k=1}^T (\Delta \mathbf{\phi}(t_k) - \overline{\Delta \mathbf{\phi}})(\Delta \mathbf{\phi}(t_k) - \overline{\Delta \mathbf{\phi}})^T \label{eq:cov_matrix}\\
    \overline{\Delta \mathbf{\phi}} &= \frac{1}{T} \sum_{k=1}^T \Delta \mathbf{\phi}(t_k) \\
    \Delta \mathbf{\phi}(t_k) &= \mathbf{\phi}(t_k) - \mathbf{\phi}(t_{k-1})
\end{align}
where $\overline{\Delta \mathbf{\phi}}$ is the sample mean. $\Delta \mathbf{\phi}(t_k)$ is the N-dimensional vector of differential comb line phase noise at time $t_k$. 

The correlation matrices $\mathbf{R}$ are then calculated component-wise from the covariance matrices using
\begin{align}
    \mathbf{R}_{nm} = \frac{\mathbf{Q}_{nm}} {\sqrt{\mathbf{Q}_{nn}\mathbf{Q}_{mm}}}
    \label{eq:corr_matrix}
\end{align}
which returns the Pearson correlation coefficients.

In \cref{fig:correlation_matrices}, the correlation matrix estimates for the different phase estimation methods are shown. Both correlation matrices from the EKF-based variants are similar to the true matrix. In the unconstrained case, however, the correlations for the line with index $l=7$ deviate from the ground truth. This line has a very low SNR, which causes problems in the optimization process. The optimized parameters that correspond to this line are harder to find, which causes the inaccuracy.

The conventional phase estimations' correlations match the true correlations for the central lines. As discussed before, the method fails for low SNR lines, resulting in completely uncorrelated estimates for the outer comb lines.
In summary, the accuracy of the correlation matrix estimated seems to follow the phase noise variance estimate that is shown in \cref{fig:eo_comb_lines_variance}. This is not surprising as inaccurate phase noise estimates will lead to inaccurate estimates of their correlations.

\begin{figure*}
    \centering
    \includegraphics[width=\linewidth]{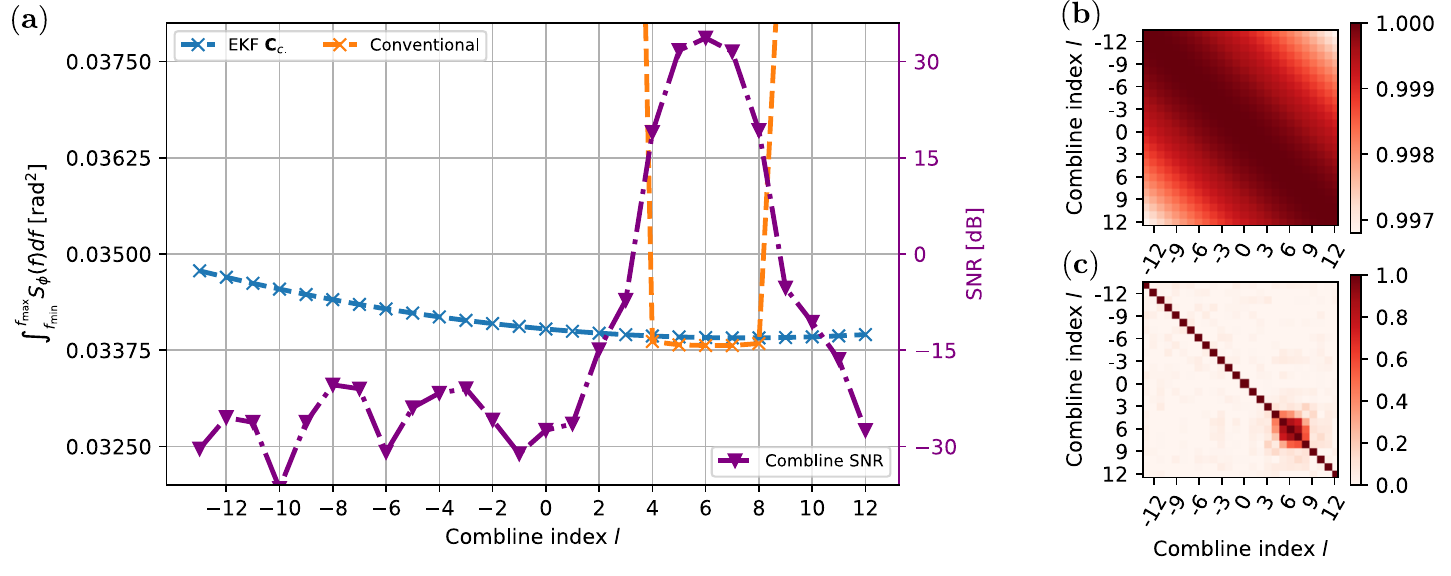}
    \caption{
    Accuracy comparison between the EKF-based and conventional phase noise estimation method on an experimental EO comb.
    (a) Integrated phase noise PSD per comb line for each method. For the EKF, the constrained model was used which only allows common and differential mode phase noise.
    Additionally, the SNR per comb line is shown with the purple, triangular markers which are measured on the right y-axis. The SNR is calculated as in Eq. \eqref{eq:SNR}.
    (b) Differential phase noise correlation matrix of the EKF-based estimate.
    (c) Differential phase noise correlation matrix of the conventional method's estimate.
    }
    \label{fig:exp_eo_variance_and_corr}
\end{figure*}

\subsection{Application to an experimental EO comb}
In this section, we demonstrate the robustness of the method and its applicability to real-world experimental data. The EKF-based method will be applied on experimental traces of a down-converted EO comb with similar characteristics to the simulated comb discussed in \cref{sec:EO_comb_sim}.

The experimental setup for the measurement considered in this section is identical to the setup displayed in \cref{fig:downconverted_comb_sim_signal} (a). A \textit{NKT Photonics Koheras BASIK E15} fiber laser is mixed with an EO comb. The EO comb is generated using a separate \textit{NKT Photonics Koheras BASIK E15} laser and an EO modulator, driven by an \textit{Agilent E8247C} signal generator. The mixed signal is detected by a balanced photodetector and digitized using a sampling oscilloscope with $f_{BW}=\SI{13}{\giga\hertz}$ and $f_{sampling}=\SI{40}{\giga\hertz}$. After digitization, the recorded signal is downsampled digitally by a factor \num{3}, as no beat notes appear in the high-frequency part of the signal spectrum. This step reduces the bandwidth of the measurement which increases the SNR.
The spectrum of the resulting measured signal is displayed in Fig. \ref{fig:exp_eo_comb_signal_and_latent_psds} (a). We identify \num{13} non-aliased down-converted comb lines and \num{13} aliased lines, which are centered around the line with index $l_{center}=6$.
Notably, all lines except lines four to eight have low powers and, therefore, low SNRs. Additionally, each aliased line happens to be only about \SI{36}{\mega\hertz} away from a neighboring non-aliased line. Since most of the regular lines have higher SNR, they dominate over the aliased lines which makes them hard to characterize.
This is a problem when using the conventional method since it relies on bandpass filtering around each line. Because of the narrow spacing, the maximum bandwidth of each filter is limited to these \SI{36}{\mega\hertz}. This limits the maximum measurable phase noise PSD offset frequency to half the bandwidth. For this measurement, this means that the conventional method can be used to characterize each comb line only up to \SI{18}{\mega\hertz}. While in practice higher bandwidths could be used, the resulting phase noise estimate would be biased since it would include an aliased line.

This issue could be mitigated in this particular experimental setup by adjusting the central frequency of the seed laser. Nevertheless, in frequency combs with low repetition rates, the narrow spacing between lines remains an inherent challenge. Consequently, this measurement highlights the benefit of employing the EKF, as it eliminates the need for band-pass filtering and therefore circumventing the problem.

As long as the state space model correctly describes the detected signal, it does not matter how close the lines are to each other in frequency. We can, therefore, separate the contribution of each line and, in principle, isolate each line from the signal.
However, here we are interested in their phase noise properties.

To apply the EKF to the detected signal, we use the same steps as in \cref{sec:EO_comb_sim}. The same state-space model as defined in Eqs. \eqref{eq:latent_model_comb_transition} and \eqref{eq:latent_model_comb_measurement} are used, which describe the latent space phase noise model.
As shown in the aforementioned section, the constrained model variant produces a more accurate phase noise estimate, especially for very low SNR comb lines. In this section we will, therefore, only use this variant. The corresponding optimization transformation function $\Lambda$ is defined in Eq. \eqref{eq:optimization_function_sim_eo_comb}.
After optimizing the model parameters $\pmb{\theta}$, the EKF is applied to estimate the phase. As we used the constrained model for the mixing matrix $\mathbf{C}_{c.}$, the latent phase noise sources can be identified as the common and differential mode noise sources.
Here the common mode corresponds to the difference of the phase noise of the comb seed laser and the LO $\phi^{CW}(t_k) - \phi^{LO}(t_k)$. Both contribute equally to all down-converted comb lines, as in Eq. \eqref{eq:EO_comb_phase_noise}.
The differential mode is the phase noise originating from the RF source $\phi^{RF}(t_k)$.

Both the common and differential mode contributions are separately accessible in the experiment, such that we can record independent phase noise reference measurements.
The common mode phase noise is estimated by creating a single beat note signal between the seed laser and the LO, which has the same total phase noise as the common mode of the EO comb.
For the RF source reference phase noise measurement, the source is connected directly to the ADC. The phase noise of both reference measurements is estimated using the conventional method.

The frequency noise PSDs of the estimated latent phase noise sources are shown in 
\cref{fig:exp_eo_comb_signal_and_latent_psds} (b), alongside the reference measurements.
The PSDs of the EKF estimates are accurate over a wide frequency range, as they match the reference measurements very well.
The common mode latent phase noise $\psi_0$ matches the seed laser phase noise over the full frequency range.
Up to \SI{1}{\mega\hertz} offset frequency, the differential mode phase noise $\psi_1$ matches the RF source phase noise.
Beyond \SI{1}{\mega\hertz}, the reference measurement is limited by measurement noise, which can be seen by the $\propto f^2$ dependency of the frequency noise PSD.
At the same frequency, the PSD of $\psi_1$ flattens out and its CI grows larger.
Still, the interval predicts the PSD to be significantly below the noise floor of the reference measurement. A flat frequency noise PSD corresponds to the random walk behavior that is part of the EKF state space model in Eq. \eqref{eq:latent_model_comb_transition}. It is, therefore, not surprising that the EKF PSD estimate takes on this behavior: Once the phase noise is limited by measurement noise, extrapolating from the state-space model is the only remaining information the EKF can use.
The CI does not reflect this model assumption. Its calculation, as presented in the appendix, does not include uncertainty in the parameters of the model itself. It represents uncertainty originating only from the measurement noise while assuming completely certainty about the state-space model.
While a random walk is a valid assumption for laser phase noise, this is not necessarily the case for an RF source, as discussed in \cref{sec:EO_comb_sim}.
The latent phase noise PSDs below the measurement noise floor should only be trusted if the phase noise behavior is well-modeled in the state-space model.
Since we can not assume that the random walk is a good model for the RF source phase noise, we can not trust the PSD of $\psi_1$ beyond \SI{1}{\mega\hertz}.

Next, we investigate the estimation accuracy with respect to the phase noise of the individual comb lines. Just as for the simulated EO comb, the phase noise per comb line can be calculated using Eq. \eqref{eq:latent_space_eq}.
They are compared to estimation using the conventional method.

\cref{fig:exp_eo_variance_and_corr} (a) shows the integrated phase noise variance per comb line for the EKF and conventional method.
The phase noise variance per comb line is calculated by integrating the phase noise PSDs of each comb line between $f_{min}=\SI{4}{\kilo\hertz}$ and $f_{max}=\SI{15}{\mega\hertz}$.
For the central comb lines with SNR > \SI{15}{\decibel}, both methods agree on the phase noise variance.
As we saw for the simulated EO comb in \cref{fig:eo_comb_lines_variance}, both methods produce accurate estimates in high SNR regimes. Therefore, we can assume that both methods produce an accurate estimate for lines four to eight.
Beyond these lines the conventional method fails. Since we have used the constrained EKF model, its estimate follows the theoretical phase noise model of the EO comb in Eq. \eqref{eq:EO_comb_phase_noise}. Just as in the simulation, the variance follows the expected parabolic form.
Since this is an experimental measurement, we do not have access to ground truth values. The simulation results from the previous section, however, indicate an accurate prediction.

In \cref{fig:exp_eo_variance_and_corr} (b,c), the estimated correlation matrices of both methods are shown, which are calculated according to Eqs. \eqref{eq:cov_matrix} and \eqref{eq:corr_matrix}.
As expected, the EKF predicts strongly correlated comb line phase noise. The conventional method can only recover the correlations for the few central lines that exhibit enough SNR.

To summarize this section, we have shown how to use the EKF to characterize the phase noise of down-converted frequency combs using the example of an EO comb.
Both in simulation and on experimental data we demonstrate that the EKF outperforms the conventional DSP method.
Especially for very low SNR comb lines, the EKF can still recover the phase noise while the conventional method fails. While the EO comb was used as a benchmark system, the latent space EKF method is much more general. Two variants of the latent space were introduced, which, in principle, allows for its application on any comb.

\section{Dual-comb setups}
\label{sec:04_dual_comb}

In this section, we investigate the dual-comb setup. We continue to utilize the same experimental setup as displayed in Fig. \ref{fig:setup}. Both the CUT and the LO are chosen to be frequency combs. In this case, the detected signal has the most general form as derived in Eq. \eqref{eq:BPD_no_rin}. Each optical comb line of the CUT can, in principle, create a beat note with any of the optical comb lines of the RS.
The number of beat notes in the detected signal depends on the repetition frequencies $\omega^{CUT}_{rep}$ and $\omega^{LO}_{rep}$ of the comb, as well as the available detection bandwidth.

The beat notes in dual-comb setups can be categorized into different orders. One line of each optical comb is arbitrarily chosen to have comb line index zero. The electrical beat note created by their mixing is a zeroth order beat note and is assigned a line index from each generating line.
Its frequency will therefore be denoted as $\omega_{0,0}=\omega^{LO}_0 - \omega^{CUT}_0$.
All beat notes generated from the neighboring comb lines of each comb with matching index are also considered zeroth order.
The index difference of the generating lines determines the order number of a beat note. A beat note with frequency $\omega_{0,1}=\omega^{LO}_0 - \omega^{CUT}_1$ would therefore be considered a beat note of order one.

\begin{figure}
    \centering
    \includegraphics[width=0.95\linewidth]{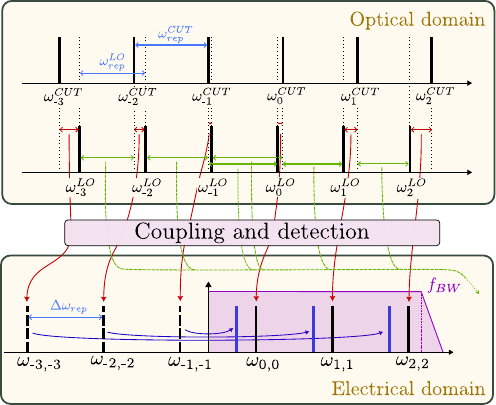}
    \caption{
    Illustration of the beat note generation process in the dual-comb setup.
    The upper panel schematically shows the comb lines of the CUT and LO in the optical domain. They are shown with slightly different repetition frequencies $\omega^{CUT}_{rep}$ and $\omega^{LO}_{rep} \equiv \omega^{CUT}_{rep} - \Delta\omega_{rep}$. Through coupling and detection, the electrical beat notes, displayed in the lower panel, are generated.
    The generation processes of all zeroth order beat notes are illustrated with red arrows.
    The green dashed arrows indicate the mixing process leading to beat notes of the order \num{-1}. In this illustration we assume that the frequencies of all higher-order beat notes lie outside the detection bandwidth $f_{BW}$, which is indicated in purple.
    Finally, the blue arrows indicate the aliasing of negative difference frequencies to positive detected frequencies.
    }
    \label{fig:dual_comb_notes_origin}
\end{figure}

In \cref{fig:dual_comb_notes_origin} the beat note generation process is illustrated. The spectrum of the zeroth-order beat notes is again a comb in the electrical domain with repetition frequency $\Delta\omega_{rep} = \omega^{CUT}_{rep} - \omega^{LO}_{rep}$.

In the illustration, we have assumed that all higher-order beat notes do not contribute to the detected spectrum.
If the repetition rates of the optical combs are tunable, it is possible to control the frequencies of the resulting beat notes. This allows us to choose which beat notes are within the detectable bandwidth. To ensure that only zeroth order notes are detected, the combs should fulfill the inequations
\begin{align}
\begin{split}
    \omega^{CUT}_{rep} &> 2\pi f_{BW} \\
    \omega^{LO}_{rep} &> 2\pi f_{BW}
    \end{split} \label{eq:dual_comb_cond1}
\end{align}
and
\begin{align}
\begin{split}
    \Delta\omega_{rep} &\ll \omega^{CUT}_{rep}\\
    \Delta\omega_{rep} &\ll \omega^{LO}_{rep}.
    \end{split} \label{eq:dual_comb_cond2}
\end{align}
The inequations (\ref{eq:dual_comb_cond1},\ref{eq:dual_comb_cond2}) are, in principle, achievable for all types of combs with sufficiently tunable repetition frequencies. By tuning the difference frequency $\Delta\omega_{rep}$ to be very small compared to the optical comb repetition frequencies (Eq. \eqref{eq:dual_comb_cond2}), the detection bandwidth can be reduced via low-pass filtering such that the first set of inequations (\ref{eq:dual_comb_cond1}) are fulfilled.

In this regime, the frequencies of all higher-order beat notes are larger than $f_{BW}$ and will not be part of the detected signal. The analytical form of the detected signal in Eq. \eqref{eq:BPD_no_rin} is reduced to
\begin{align}
\begin{split}
        y(t_k) &= 2\eta \sum_{l=-\infty}^\infty \sqrt{P^{CUT}_l P_l^{LO}} \\
        &\quad \times \text{sin}[(\omega^{CUT}_l - \omega^{LO}_{l})t_k \\
        &\qquad  + \phi^{CUT}_l(t_k) - \phi^{LO}_l(t_k)] + \xi(t_k)\\
         &\equiv 2\eta \sum_{l=-\infty}^\infty A_{l} \text{sin}[\omega_{l,l} t_k + \phi^{total}_{l,l}(t_k)] + \xi(t_k)
         \end{split} \label{eq:dual_comb_signal}
\end{align}
where we only consider zeroth order terms. As no higher-order terms appear, we do not need a double summation in the expression.
Note that here we used the index notation $l$ as presented in Fig. \ref{fig:dual_comb_notes_origin}, where the comb line corresponding to $l=\num{0}$ is arbitrarily chosen for each comb. 
The beat note frequencies are 
\begin{align}
\begin{split}
    \omega_{l,l} &= \omega^{CUT}_l - \omega^{LO}_{l} \\
    &= \underbrace{\omega^{CUT}_{0} - \omega^{LO}_{0}}_{\omega_{0,0}} + l \underbrace{\left( \omega^{CUT}_{rep} - \omega^{LO}_{rep} \right)}_{\Delta\omega_{rep}} \\
    &= \omega_{0,0} + l\Delta\omega_{rep}.
\end{split} \label{eq:dual_comb_beat_freqs}
\end{align}
The phase noise terms $\phi^{total}_{l,l}$ are the sums of the corresponding phase noise terms of the optical combs. 
Notably, the signal of the dual-comb setup in Eq. \eqref{eq:dual_comb_signal} now has the same form as the signal from the down-converted comb setup in Eq. \eqref{eq:downconverted_comb_signal}. Since the analytical form is identical we can use the same state-space model for the EKF-based estimation method utilizing the latent phase noise space. The state-space model is defined in Eqs. \eqref{eq:latent_model_comb_transition} and \eqref{eq:latent_model_comb_measurement}.

\subsection{Experimental dual mode-locked laser}
\begin{figure*}
    \centering
    \includegraphics[width=\linewidth]{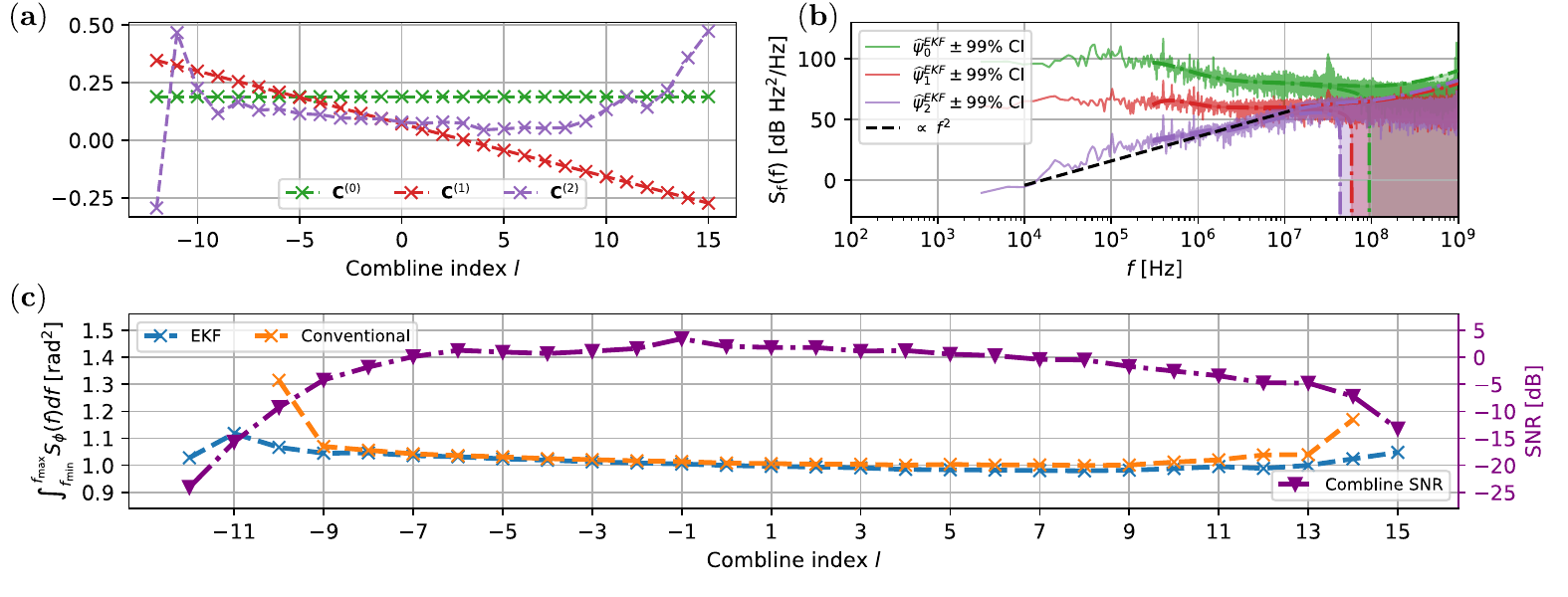}
    \caption{EKF-based phase noise characterization in an experimental dual-comb MLL setup.
    (a) Column vectors of the optimized latent space mixing matrix \textbf{C}, indicated by the superscripts $^{(0,1,2)}$.
    (b) Frequency noise PSDs of the latent phase noise sources $\psi_l$ as tracked by the EKF. The shaded areas represent the 99\% CIs of the PSDs, which were smoothed for visibility.
    (c) Integrated phase noise PSD per comb lines.
    The blue and orange markers represent the variance of the EKF estimate and the conventional method respectively, measured on the left y-axis. The purple triangular markers show the SNR per comb line, measured on the right y-axis.
    The integration limits are $f_{min}=\SI{5}{\mega\hertz}$ and $f_{max}=\SI{210}{\mega\hertz}$.}
    \label{fig:dual_comb_exp_all}
\end{figure*}

We will demonstrate the application of the EKF in dual-comb setups using an experimental measurement with two frequency-modulated quantum dot mode-locked lasers (MLL).
A description of the fabrication and a general characterization of the MLLs used in the measurement are given in \cite{dumont_high-efficiency_2022}.
Both MLLs were fabricated in the same way and should have similar noise properties. The distinction between CUT and LO is, therefore, not useful. We will, however, keep the naming convention to maintain the notation.
Their repetition frequencies are $\omega^{CUT}_{rep}\approx\SI{59.4}{\giga\hertz}$ and $\omega^{LO}_{rep}\approx\SI{58.4}{\giga\hertz}$. Both combs are centered around $\lambda=\SI{1298}{\nano\meter}$ with an average comb line power of $\SI{-20}{\dBm}$.
The beat notes have a spacing of $\Delta\omega_{rep}\approx\SI{1}{\giga\hertz}$. 
A single photodetector with \SI{15}{\giga\hertz} bandwidth was used to detect the signal. The resulting spectrum contains 16 regular and 12 aliased comb lines. With these parameters, the inequations (\ref{eq:dual_comb_cond1}, \ref{eq:dual_comb_cond2}) are fulfilled. The signal spectrum therefore only contains zeroth-order beat notes.

While we can use the same state-space model as for the EO comb, we have to decide whether the mixing matrix that defines the latent phase noise space should be constrained.
In \cref{sec:03_downconverted_combs} we used an optimization transformation function $\Lambda$ (as defined in Eq. \eqref{eq:optimization_function_sim_eo_comb}) which constrains the latent phase noise sources. In particular, we chose to constrain $\mathbf{C}_{c.}$ to two latent noise sources: a common mode, and a differential mode source. This model was chosen as in the EO comb, a comb line index-dependent model of the phase noise is known and defined in Eq. \eqref{eq:EO_comb_phase_noise}.

For the MLLs used in this experiment, we do not have an explicit phase noise model available.
However, we can use the so-called elastic tape model for phase noise to our advantage. It assumes that the physical noise sources that contribute to the comb line phase noise have a linear dependence with the comb line number \cite{benkler_circumvention_2005}. Each physical noise source may, however, differ in their fixed point, which is the comb line index at which its contribution is minimal.
This model has been shown to describe the comb line phase noise dependency in MLLs as well as microring-resonator combs \cite{lei_optical_2022, hutter_femtosecond_2023}.

This is equivalent to the assumption that any phase noise sources that are present in the comb behave as differential mode phase noise sources which are centered at different comb lines.
This relationship can be expressed as
\begin{align}
    \phi^{CUT}_l(t) = \sum_n (l - l_{fix, n})\varphi_n(t)
\end{align}
where $\varphi_n(t)$ is a physical noise source centered at the comb line with index $l_{fix, n}$.
The total phase noise per comb line is then the sum of the independent physical noise source contributions.
As the contribution of each source is linear with respect to the comb line index $l$, the total phase noise dependence on line number must also be linear.
If there are at least two physical noise sources with different fixed points, the total phase noise may also have a line number independent common mode contribution.
This means that the total phase noise in each comb can be described by the decomposition into one common mode and one differential mode latent phase noise source.
By the same logic, the phase noise in the electrical beat notes can also be described by only two noise modes.

This is the same decomposition we have used in the EO comb case.
We could, therefore, use the same two-dimensional latent space constraints as before. To, however, showcase the versatility of the method and the validity of the assumptions above, we propose a hybrid latent space model. We choose the latent space dimension $L=\num{3}$, where the first two noise sources are constrained to common and differential mode. Analogous to the EO comb, we constrain the first column of \textbf{C} to be constant with line index and the second to be linear, with an optimizable offset $l_{center}$. These two latent dimensions represent the common mode and differential mode noise sources.
The third column of \textbf{C} is left unconstrained. If any additional noise sources are present that do not follow the elastic tape model, this additional latent space component should capture their behavior.

To improve the convergence behavior, we have found that normalizing the columns of \textbf{C} drastically improves the robustness and stability of the optimization. As we are not interested in interpreting the latent noise sources in this case, changing the scale of \textbf{C} is not a problem. The normalization is applied after the optimization transformation function as
\begin{align}
    \mathbf{C}^{(m)} \rightarrow \mathbf{C}^{(m)} / ||\mathbf{C}^{(m)}||_2
\end{align}
for the m-th column vector of $\mathbf{C}$ where $||\cdot||_2$ is the 2-norm.
Apart from this change, the same optimization transform function $\Lambda$ as in Eq. \eqref{eq:optimization_function_sim_eo_comb} is used.

While we can not interpret the latent space noise sources as physical quantities, we can still gain insights by investigating their properties. In Fig. \ref{fig:dual_comb_exp_all} (a,b) the column vectors of the optimized \textbf{C} matrix are shown, alongside the frequency noise PSDs of the latent sources. 
The two constrained vectors $\mathbf{C}^{(0)}$ and $\mathbf{C}^{(1)}$ look as expected, where the optimized offset of the zero-crossing of the differential mode vector occurs at $l_{center}^{MAP} = \num{3}$.
The third free column of \textbf{C} exhibits a mostly smooth shape with smaller values at the central lines and larger values towards the outer lines.
The frequency noise PSDs of the latent sources reveal the true scale of each noise mode. The common mode noise is the largest latent source and is the dominant contribution to the phase noise per comb line.
The differential mode is significantly smaller and has a mostly flat frequency noise PSD profile.

The free noise mode $\psi_3$ has an approximate $\propto f^2$ dependency on the PSD offset frequency. This makes the third latent noise source a white noise contribution to the comb line phase noise.
it very likely that some of the measurement noise present in the signal is interpreted as phase noise by the EKF.
While this is generally undesirable, it also shows that during the optimization process, the EKF was not able to find phase noise contributions that violate the elastic tape model. This is not proof of the non-existence of higher-order noise terms, but it does show that the comb line phase noise can be accurately estimated using only the common and differential noise mode model.

Finally, Fig. \ref{fig:dual_comb_exp_all} (c) displays the intergrated phase noise PSD per comb line. The EKF estimate is compared to the conventional method. As we have seen before, both methods agree for the high SNR comb lines. For the outer low SNR lines, the conventional method returns a significantly larger variance estimate compared to the EKF.

\subsection{Digital noise compensation}
We have shown how the EKF can be used to characterize the total phase noise of frequency combs.
However, the EKF can also be used to digitally compensate for the phase and measurement noise of a multi-heterodyne signal.

As the distinction between the actual signal and perturbing measurement noise is inherent to the method, filtering out the measurement noise is straightforward. After optimizing the parameters $\pmb{\theta}$ of the state-space model and applying the EKF, we are left with a time series of the latent phase noise estimates $\psi_m(t_k)$. 
If we iterate over the samples $k$ and use $\psi_m(t_k)$ as an input to the state-space model, we can therefore reconstruct the estimate of the EKF for the signal $\widetilde{y}^{EKF}(t_k)$ without the measurement noise $\xi(t_k)$, such that $y(t_k) = \widetilde{y}^{EKF}(t_k) - \xi(t_k)$.

In principle, the phase noise of the detected comb lines can be compensated for in the same way. Using the latent space Eq. \eqref{eq:latent_space_eq}, we receive a phase noise time series $\phi_l(t_k)$ for each comb line. Applying the inverted phase noise $-\phi_l(t_k)$ as a phase shift to each line in $\widetilde{y}^{EKF}(t_k)$ should result in a phase and measurement noise-free estimate of the original noisy signal $y(t_k)$.

In a similar approach, it has been shown that the EKF can be used to recover comb lines in significantly more noisy conditions \cite{burghoff_generalized_2019, burghoff_computational_2016}. In dual-comb spectroscopy experiments, it is common to produce very narrowly spaced multi-heterodyne signals to achieve good frequency resolution. If the combs have significant phase noise, the fluctuations of the comb line frequencies can cause overlap in the spectrum. Even in situations where individual lines were not discernible in the spectrum, using the EKF to compensate for the phase noise was possible.

Digital noise compensation methods can relax the requirements of frequency combs for metrology applications \cite{sterczewski_computational_2019}. Instead of relying on complex experimental setups to lock combs, the phase noise can be removed retroactively. While the presented method involves complex DSP and requires significant computational processing power, further improvements to its efficiency could enable the use of free-running combs in dual-comb spectroscopy applications.

\section{Conclusion}
\label{sec:conclusion}
In this tutorial, we have shown how to characterize laser and frequency comb phase noise using multi-heterodyne measurements and the extended Kalman filter. By applying automatic differentiation and adaptive optimization methods, a state space model of the detected multi-heterodyne signal with arbitrary shape can be learned from a measured signal. On simulated and experimental datasets in different experimental setups, we have demonstrated the method's versatility and its advantages over conventional DSP approaches.
Further we have discussed how the framework can be used for offline digital phase and measurement noise compensation.
Wideband noise compensation in free-running dual-comb spectroscopy could reduce the hardware requirements on the combs and eventually lead to more efficient operation.

\begin{acknowledgments}
This work has been funded by the SPOC Centre (DNRF 123) and Villum Fonden (VI-POPCOM 54486). The authors thank Mario Dummont, Osama Terra, Bozhang Dong and John E. Bowers from University of Carlifornia, Santa Barbara, USA for providing the dual-comb MLL data.
\end{acknowledgments}

\section*{Conflicts of interest and data availability}
The authors have no conflict of interest to disclose. The data and code supporting the findings in this work are available from the corresponding author upon reasonable request.

\appendix*
\section{Second-order uncertainty propagation to calculate confidence intervals of power spectral densities}
\label{sec:appendix}
In this section, we derive an expression for the confidence interval of a power spectral density given a sampled time series with time-dependent uncertainties.
In our case, the expression is used to propagate the EKF uncertainty estimates of the phase noise time series to the phase noise PSD. This allows the quantification of the EKF's uncertainty in the PSD picture.
The approach is similar to a derivation presented in \citep{palczynska2005uncertainty}. Our derivation is, however, generalized for time-dependent uncertainties and exact due to the higher-order error propagation.

\paragraph{PSD estimation}
To estimate the PSD of a time series $x(t_k)$, Welch's method is used \citep{welch_use_1967}. $t_k=k\Delta T$ is the discretized time variable with sampling period $\Delta T$.
$x(t_k)$ is divided into $N_{seg.}$ potentially overlapping segments of length $K$, and each segment is multiplied by a window function $w(k)$ of the same length. Using
\begin{subequations}
\begin{align}
    S_n(l) &= \frac{2 \Delta T}{W} |X(l)_n|^2 \nonumber \\
           &= \frac{2 \Delta T}{W} \left( \mathrm{Re}\left[X(l)_n\right]^2 + \mathrm{Im}\left[X(l)_n\right]^2 \right) \label{eq:segment_psd}\\
    W &= \sum_{k=0}^{K-1} |w(k)|^2
\end{align}
\end{subequations}
a periodogram is calculated for each segment. The index $l$ corresponds to the Fourier frequency $f_l = l / (K \Delta T)$. $X(l)_n$ is the discrete Fourier transform of the n-th windowed segment, defined as 
\begin{subequations}
\begin{align}
    X(l) &= \mathrm{Re}[X(l)] + i\ \mathrm{Im}[X(l)] \\
    \mathrm{Re}[X(l)] &= \sum_{k=0}^{K-1} w(k) x(t_k) \cos{\left[-2\pi kl/K\right]}\\
    \mathrm{Im}[X(l)] &= \sum_{k=0}^{K-1} w(k) x(t_k) \sin{\left[-2\pi kl/K\right]}
\end{align}
\label{eq:fourier}
\end{subequations}

After computing $S_n(l)$, these periodograms are averaged to obtain the final PSD estimate using
\begin{align}
    S(l) = \nicefrac{1}{N_{seg.}} \sum_{n=0}^{N_{seg.} - 1} S_n(l).
\end{align}

\paragraph{Uncertainty propagation}
We assume we are given a time series $u_x(t_k)$ of the uncertainty of $x(t_k)$. In our case, $u_x(t_k)$ is the standard deviation of the Gaussian distribution the EKF predicts.

In the calculation of $S(l)$, $x(t_k)$ first enters in the calculation of the Fourier transform in Eq. \eqref{eq:fourier}. To propagate the uncertainty, the real and imaginary parts of $X(l)$ are treated separately \citep{betta_propagation_2000}.
As the Fourier transform is linear in $x(t_k)$, we can use linear error propagation:
\begin{subequations}
    \begin{align}
        u^2_{\mathrm{Re}[X_l]}(l) &= \sum_{k=0}^{K-1} \left( \frac{\partial \mathrm{Re}[X_l]}{\partial x(t_k)} \right)^2 u_x^2(t_k) \\
        &= \sum_{k=0}^{K-1} w(k)^2 \cos{\left( 2\pi k l / K \right)}^2 u_x^2(t_k) \nonumber \\
        &= \sum_{k=0}^{K-1} w(k)^2 \frac{1 + \cos{\left( 4\pi k l / K \right)}}{2} u_x^2(t_k) \nonumber  \\
        &= \sum_{k=0}^{K-1} w(k)^2 u_x^2(t_k)/2 \nonumber \\
         &\qquad + \mathrm{Re}\left[ \sum_{k=0}^{K-1} e^{\left( 2\pi (2k) l / K \right)} w(k)^2 u_x^2(t_k)/2 \right] \label{eq:fft_prop_re}
    \end{align}
\end{subequations}
We have reformulated the second term of Eq. \eqref{eq:fft_prop_re} to showcase that this expression again is a discrete Fourier transform.
Analogously, the imaginary part is given by
\begin{align}
\begin{split}
    u^2_{\mathrm{Im}[X_l]}(l) &= \sum_{k=0}^{K-1} w(k)^2 u_x^2(t_k)/2 \\
         &\qquad - \mathrm{Re}\left[ \sum_{k=0}^{K-1} e^{\left( 2\pi (2k) l / K \right)} w(k)^2 u_x^2(t_k)/2 \right]. \label{eq:fft_prop_im}
 \end{split}
\end{align}

Higher-order error propagation is required to calculate the uncertainty of $S_n(l)$ exactly since Eq. \eqref{eq:segment_psd} is not linear. The expression is quadratic, which means that second-order error propagation is sufficient as all higher-order sensitivity coefficients vanish.
Using the analytical form given in \citep{mekid_propagation_2008}, we arrive at the expression
\begin{align}
\begin{split}
    u^2_{S_n}(l) &= \left(\frac{\partial S_n}{\partial (\mathrm{Re}[X_l])}\right)^2 u^2_{\mathrm{Re}[X_l]}(l) \\
    &\quad + \frac{\kappa-1}{4} \left( \frac{\partial^2 S_n}{\partial (\mathrm{Re}[X_l])^2}\right)^2 u^4_{\mathrm{Re}[X_l]}(l) \\
    &\quad + \left(\frac{\partial S_n}{\partial (\mathrm{Im}[X_l])}\right)^2 u^2_{\mathrm{Im}[X_l]}(l) \\
    &\quad + \frac{\kappa-1}{4} \left(\frac{\partial^2 S_n}{\partial (\mathrm{Im}[X_l])^2} \right)^2 u^4_{\mathrm{Im}[X_l]}(l). \label{eq:u_Sn}
\end{split}
\end{align}
As the EKF produces normally distributed estimates, we will use $\kappa=\num{3}$, the Kurtosis of a Gaussian distribution. 

The sensitivity coefficients evaluate to
\begin{subequations}
\begin{align}
    \frac{\partial S_n}{\partial (\mathrm{Re}[X_l])} &= \frac{4\Delta T}{W} \mathrm{Re}[X_l] \\
    \frac{\partial^2 S_n}{\partial (\mathrm{Re}[X_l])^2} &= \frac{4\Delta T}{W} \\
    \frac{\partial S_n}{\partial (\mathrm{Im}[X_l])} &= \frac{4\Delta T}{W} \mathrm{Im}[X_l] \\
    \frac{\partial^2 S_n}{\partial (\mathrm{Im}[X_l])^2} &= \frac{4\Delta T}{W}.
\end{align}
\end{subequations}
Substituting into Eq. \eqref{eq:u_Sn} yields
\begin{align}
\begin{split}
    u^2_{S_n}(l) &= \frac{16\;\Delta T^2}{W^2} \left( \mathrm{Re}[X_l]^2 u^2_{\mathrm{Re}[X_l]}(l) + \mathrm{Im}[X_l]^2 u^2_{\mathrm{Im}[X_l]}(l) \right. \\
    &\qquad \left. + \frac{1}{2} u^4_{\mathrm{Re}[X_l]}(l) +\frac{1}{2} u^4_{\mathrm{Im}[X_l]}(l)\right).
\end{split}
\end{align}

Finally, the uncertainty of the averaged PSD estimate $S(l)$ can be calculated according to linear uncertainty propagation:
\begin{align}
\begin{split}
    u^2_{S}(l) &= \sum_{n=0}^{N_{seg.}-1} (\frac{\partial S}{\partial S_n})^2 u^2_{S_n}(l) \\
    &= \nicefrac{1}{N_{seg.}} \sum_{n=0}^{N_{seg.}-1} u^2_{S_n}(l).
\end{split}
\end{align}
This expression is used throughout the paper to calculate the confidence intervals of PSDs, as shown in the figures.

\end{document}